\documentclass[prd,twocolumn,nofootinbib,superscriptaddress,amsmath,amssymb,eqsecnum]{revtex4-1}
\usepackage[utf8]{inputenc}
\usepackage[T1]{fontenc}
\usepackage{mathtools}
\usepackage{graphics}
\usepackage{graphicx}
\usepackage{dcolumn}
\usepackage{bm}
\usepackage{dsfont} 
\usepackage{amsmath,amssymb}
\usepackage{hyperref}
\usepackage{tabularx}
\usepackage{epstopdf}
\usepackage{tensor}
\usepackage{mathrsfs}
\usepackage[normalem]{ulem}
\usepackage[usenames]{color}
\hypersetup{
    colorlinks=true,
    linkcolor=blue,
    filecolor=magenta,      
    urlcolor=blue,
    citecolor=blue
}
\usepackage[dvipsnames]{xcolor}
\usepackage{mathrsfs}
\usepackage{microtype}

\newcommand{\red}[1]{\textcolor{black}{#1}}
\newcommand{\mathred}[1]{\color{black}{#1}}
\newcommand\be{\begin{equation}}
\newcommand\ba{\begin{eqnarray}}
\newcommand\ee{\end{equation}}
\newcommand\ea{\end{eqnarray}}
\newcommand\bw{\begin{widetext}}
\newcommand\ew{\end{widetext}}
\newcommand{\lb}{\left(}
\newcommand{\rb}{\right)}

\newcommand{\guelph}{Department of Physics, University of Guelph, 50 Stone Road E., Guelph, ON N1G 2W1, Canada}
\newcommand{\PI}{Perimeter Institute for Theoretical Physics, Waterloo, ON N2L 2Y5, Canada}

\begin{document}
\allowdisplaybreaks
\title{Science potential for stellar-mass black holes as neighbours of Sgr A*}

\author{Shammi Tahura}
\email{stahura@uoguelph.ca}
\affiliation{\guelph}
\affiliation{\PI}

\author{Zhen Pan}
\affiliation{\PI}

\author{Huan Yang}
\email{hyang@perimeterinstitute.ca}
\affiliation{\guelph}
\affiliation{\PI}

\date{\today}

\begin{abstract}
It has been suggested that there is possibly a class of stellar-mass black holes (BHs) residing near (distance $\le 10^3 M$) the galactic center massive black hole, Sgr A*. Possible formation scenarios  include the mass segregation of massive stellar-mass black holes and/or the disk migration if there was an active accretion flow near Sgr A* within $\mathcal{O}(10)$ Myr. In this work, we explore the application of this type of objects as sources of space-borne gravitational wave detectors, such as Laser Interferometer Space Antenna (LISA). We find it is possible to probe the spin of Sgr A* based on the precession of the orbital planes of these stellar-mass black holes moving around Sgr A*. We also show that the dynamical friction produced by accumulated cold dark matter near Sgr A* generally produces small measurable phase shift in the gravitational waveform.
In the case that there is an axion cloud near Sgr A*, the dynamical friction induced modification to gravitational waveform is measurable only if the mass of the axion field is in a narrow range of the mass spectrum. Gravitational interaction between the axion cloud and the stellar-mass black holes may introduce additional precession around the spin of Sgr A*. This additional precession rate is generally weaker than the spin-induced Lense-Thirring precession rate, but nevertheless may contaminate the spin measurement in a certain parameter regime. At last, we point out that the multi-body gravitational interaction between these stellar-mass black holes generally causes negligible phase shift during the LISA lifetime.
\end{abstract}
\maketitle

\section{Introduction}
In the past years there were exciting revolutions in probing astrophysical black holes in the strong-gravity regime, thanks to the observations by the LIGO and the Virgo collaboration (the LVC)~\cite{LIGOScientific:2021psn,LIGOScientific:2020ibl,LIGOScientific:2021usb,LIGOScientific:2018mvr} and EHT (Event Horizon Telescope)~\cite{EventHorizonTelescope:2019dse,EventHorizonTelescope:2019ggy}. In the next decades, space-based gravitational wave (GW) detectors, such as LISA~\cite{Audley:2017drz} and Tianqin~\cite{TianQin:2015yph,TianQin:2020hid}, will enable unprecedented measurements on the mass, spin, spacetime, orbital, and environmental properties of massive black holes at cosmological distances~\cite{Glampedakis:2005cf,Barack:2006pq}. For example, one particularly interesting type of source is EMRI (extreme mass ratio inspirals), which often comprises a stellar-mass black hole orbiting around a massive black hole. As this kind of system usually stays within the LISA band for $10^4-10^5$ cycles during the observation period, small additional effects (e.g., extra channel of dissipation) may accumulate over many orbital cycles to generate measurable phase shifts. As a result, EMRIs may serve as an ideal probe for deviations from General Relativity~\cite{Hannuksela:2018izj,Zhang:2019eid,Zhang:2018kib,Glampedakis:2005cf,Barack:2006pq} and astrophysical environmental effects~\cite{Bonga:2019ycj,Yang:2019iqa,Yang:2017aht,Barausse:2014tra}. In addition, the formation of EMRIs may be greatly accelerated by accretion flows, which provides promising opportunities for multi-messenger observations 
\cite{Pan2021a,Pan2021b}.

There is one special massive black hole (MBH), Sgr A*, which resides at the center of our own galaxy, with distance of orders of magnitude closer than any other massive black holes. The chance of forming an ordinary EMRI around Sgr A* during the LISA observation period is negligible ($\lesssim 10^{-6}$ from the loss-cone formation channel with EMRI rate $\sim 10^2\ {\rm Gyr}^{-1}$ per MBH \cite{Amaro2011, Babak2017,Amaro2018, Amaro2020}). However, because of the small distance between Sgr A* and earth, it is still possible to probe stellar-mass black holes at the lower end of the LISA frequency band ($\ge 10^{-5}$Hz), corresponding to orbital radius $\le 10^2 M$~\cite{Naoz:2019sjx}, where $M$ is the gravitational radius of the MBH. On the other hand, there are in fact multiple ways to produce such low-frequency EMRIs (referred as ``Sgr A* EMRI'' hereafter). For example, it has been suggested that if there is a massive subclass of stellar-mass black holes in the galactic nuclear cluster, the mass segregation effect will lead to condensation of these black holes at galactic centers, with distances $\mathcal O(10^2) M$ away from Sgr A* \cite{Emami2020,Emami2021}. In addition, if there is a previous active accretion phase in Sgr A*, as supported by the presence of stellar disk near Sgr A* \cite{Levin2003}, the disk-assisted migration tend to lead to a set of stellar-mass black holes  accumulating at distances between $\sim 80 M - 200M$ depending on the disk models and parameters \cite{Gilbaum2021,Pan2021c,Pan2021d}, as required by  equating the disk migration timescale with the gravitational wave radiation timescale. Frequencies of GWs from such Sgr A* EMRIs are within the range of $[.5-2]\times10^{-5}$ Hz, considering circular orbits. If the time that the latest accretion episode ended is not earlier than $\mathcal{O}(10^7)$ years, we still expect some black holes have not merged with Sgr A* after the gravitational wave radiation-induced inspiral, as another way of forming Sgr A* EMRIs at present. Notice that gravitational wave measurement will likely be the only way to probe Sgr A* EMRIs if they do exist.

Assuming the presence of Sgr A* EMRIs, it is natural to ask what is the detectability of these systems using LISA, and what we can learn by observing them. The first question is partially answered by previous works in~\cite{Gourgoulhon:2019iyu} by computing the event SNR (signal-to-noise ratio) as a function of orbital radius. For the sake of completeness, we  present a distribution of SNR with system parameters drawn from Monte-Carlo samplings, assuming the detection by LISA. There is one subtle point worth noticing here: the detection threshold of SNR for such systems should be much smaller than $\sim 16$, which is commonly used as a benchmark value for EMRIs. For Sgr A* EMRIs the template bank contains much less parameters as the sky location and distance of the source is known. Smaller parameter space in the waveform results in smaller penalty in the false-alarm probability, which leads to smaller detection threshold SNR $\mathred{\sim 10}$.

Our work is primarily focusing on the second question, i.e., the applications of Sgr A* EMRIs. We consider three possible sources that may produce observable effect in the gravitational waveform. The first source is the spin-orbit coupling between Sgr A* and the stellar-mass black hole, which induces Lense-Thirring precession on the orbit. We construct the waveform model to include precession effect and perform the Fisher analysis to address the accuracy of the inferred spin. The result is rather promising: for Sgr A* EMRIs with orbital separation $a \le 90 M$, we generally have $\mathred{\Delta S/S \sim 0.5\% -24\%}$, with median value at $\sim 2\%$ level. Notice that while we have measurement on the mass of Sgr A* based on the motion of S stars, the spin is completely unknown. In the future we may obtain constraints on the spin using EHT, but based on the experience from the measurement on M87~\cite{Tamburini:2019vrf,Bambi:2019tjh}, the spin constraint is unlikely to be very stringent. Measuring gravitational waves from Sgr A* EMRIs may be the most accurate way to probe the spin of Sgr A*.

Secondly, we have investigated possible influence of dark matter in the vicinity of Sgr A*. Assuming cold dark matter description with plausible density prescription at galactic center, we find that Sgr A* EMRIs are less capable of probing the effect due to dynamical friction than normal EMRIs, assuming comparable event SNR. On the other hand, if an axion dark matter cloud is excited by black hole superradiance around Sgr A*, considering a typical system configuration, we find the dynamical friction effect 
is above the measurement noise only if the mass of the axion field is between \red{$3.1\times 10^{-18}$} eV to $3.3\times 10^{-18}$ eV. The chance that the axion mass lies in such narrow range is small.  Moreover, in this case the gravitational coupling between the axion cloud and the stellar-mass BH may be strong enough to induce additional precession of the orbit plane which is larger than the measurement uncertainty. This additional precession is generally weaker than the Lense-Thirring precession, but they are degenerate with each other.

Thirdly, we also considered the influence of gravitational interaction between the stellar-mass black holes. As a model problem, we compute the Kozai-Lidov timescale assuming the presence of a nearby third body (stellar-mass black hole), and find it to be much smaller than the observation period of LISA. As a result, we conclude that multi-body effect is not important for Sgr A* EMRIs.

The rest of this article is organized as follows. In section~\ref{sec:SpinPrecession} study the effect of spin induced orbital precession to gravitational waveforms in case of Sgr A* EMRIs and determine the prospect of measuring the spin of Sgr A* with a 10 years observation period of LISA. In section~\ref{sec:dark_matter} we explore the prospect of detecting the signature of dark matter cloud from similar observations, focusing on cold dark matter in section~\ref{sec:dark_matter_cold} and axion dark matter in~\ref{sec:dark_matter_axion}. Finally, we discuss the Kozai-Lidov effect in the presence of a third object and present our concluding remarks in section~\ref{sec:discussion}. Throughout this paper, we use units in which $\hbar=c=G=1$.
\section{Spin induced orbital precession}\label{sec:SpinPrecession}
We begin by assessing the spin induced orbital precession of an Sgr A* EMRI. We will first discuss the gravitational waveform generated by such a binary and the modification to its phase due to  orbital precession in Sec.~\ref{subsec:waveform} and Sec.~\ref{subsec:precession_phase} respectively. Next, we present the data analysis formalism (Fisher analysis) that we implement for the estimation of measurement uncertainties and summarize the results in Sec.~\ref{subsec:data_results}. The main results of this section include the threshold SNR for detecting a signal from Sgr A* EMRI with LISA and constraints on various waveform parameters, including the spin of Sgr A*.
\subsection{Waveform}\label{subsec:waveform}
\begin{figure}
\begin{center} 
\includegraphics[width=8.5cm]{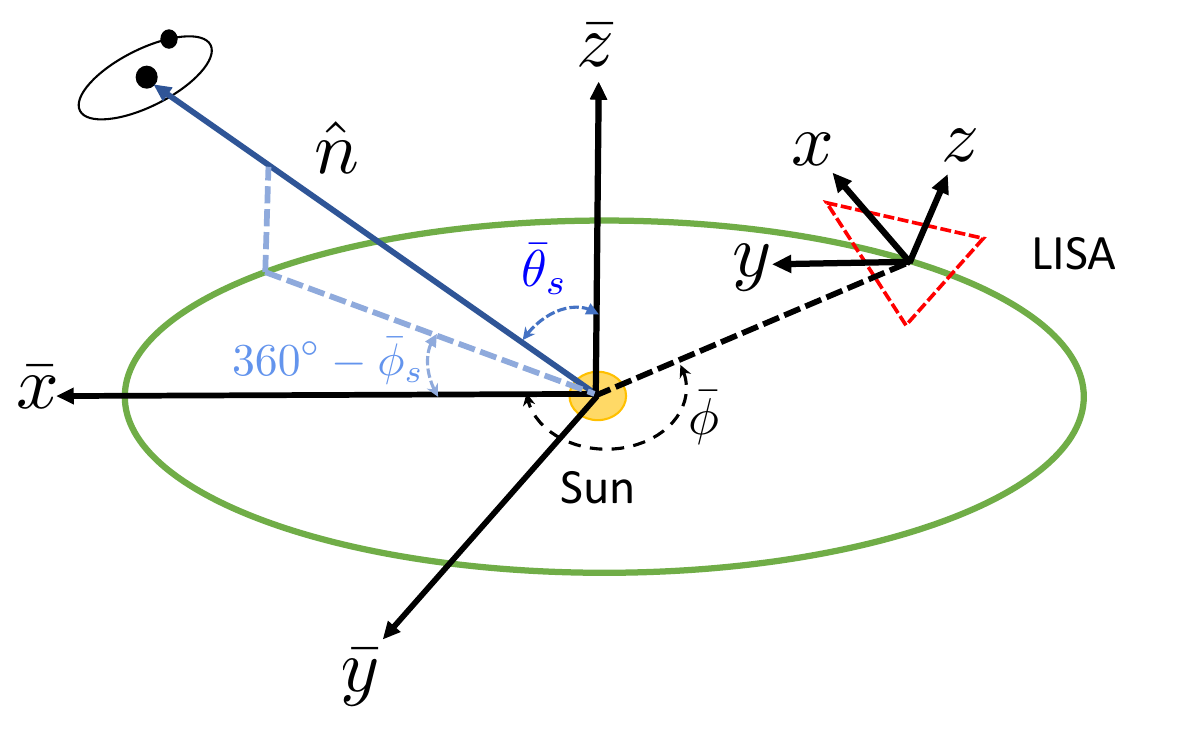}
\caption{An illustration of the coordinate systems we use in Sec.~\ref{sec:SpinPrecession}. The set of barred Cartesian coordinates $(\bar x, \bar y, \bar z)$ is tied to the eclicptic with the origin at the sun. On the other hand, the unbarred coordinates $(x, y, z)$ are tied to the detector LISA, with the detector's arms lying in the $x-y$ plane. Angle $\bar \phi (t)$ describes the location of the detector with respect to the barred coordinates in $\bar x-\bar y$ plane. $(\bar \theta_s,\bar \phi_s)$ are the spherical polar angles associated with location of the source in barred coordinates.}
\label{fig:coords}
\end{center}
\end{figure}

We implement the time domain waveform for LISA presented in Ref.~\cite{Cutler:1997ta}. Following Ref.~\cite{Cutler:1997ta}, we make use of two sets of Cartesian coordinates-- one associated with the detector denoted by $(x,y,z)$ and the other $(\bar x,\bar y,\bar z)$ tied to the ecliptic in a heliocentric frame (see Fig.~\ref{fig:coords}). The corresponding spherical polar coordinates are $(\theta, \phi)$ and $(\bar \theta,\bar \phi)$ respectively. The detector arms lie in the $x-y$ plane and the plane corotates with the detector.

The detector strain for LISA in time domain can be expressed as~\cite{Cutler:1997ta}
\ba\label{eq:waveform}
h(t)=\frac{\sqrt{3}}{2} A(t) \cos \left[2 \pi f t+\pi \dot f t^2+\varphi_{pol}(t)\right.\nonumber\\
\left.+\varphi_{D}(t)+\varphi_p(t)\right]\,.\nonumber\\
\ea
Here $A(t)$ is the amplitude, $\varphi_{pol}(t)$ is the polarization phase and $\varphi_D(t)$ is the Doppler phase that takes into account the motion of the detector with respect to the sun. We also include the effect of a small constant frequency evolution $\dot f$, which may come from various dissipation mechanisms of the binary. For the spin induced orbital precession, we set $\dot f=0$, however this term will be considered in a later part of the paper. The unconventional factor of $\sqrt{3}/2$ is attributed to the  $60^{\circ}$ angle  between the detector arms. $\varphi_p(t)$ is the modulation to  GW phase due to the orbital precession. $A(t)$ and $\varphi_{pol}(t)$ are given by
\ba
&A(t)=\left(A_{+}^{2} F_+^{2}(t)+A_{\times}^{2} F_{\times}^2(t)\right)^{1 / 2}\,, \\
&\varphi_p(t)=\tan ^{-1}\left(\frac{-A_{\times} F_{\times}(t)}{A_{+} F_{+}(t)}\right)\,,
\ea
where $F_+$ and $F_\times$ are the antenna pattern functions:
\begin{subequations}
\ba
F_{+}(t)&=&\frac{1}{2}\left(1+\cos ^{2} \theta_s\right) \cos 2 \phi_s\cos 2 \psi_s\nonumber\\
&&-\cos \theta_s \sin 2 \phi_s \sin 2 \psi_s\,,\\
F_{\times}(t)&=&\frac{1}{2}\left(1+\cos ^{2} \theta_s\right) \cos 2 \phi_s\sin 2 \psi_s\nonumber\\
&&+\cos \theta_s \sin 2 \phi_s \cos 2 \psi_s\,.
\ea
\end{subequations}
Here $(\theta_s,\phi_s)$ are polar spherical coordinates associated with $\hat n$ where we use the convention that GWs from the source propagate along $-\hat n$. Note that $(\theta_s,\phi_s)$ are time-dependent and we need to express them in terms of fixed coordinates $(\bar\theta_s,\bar\phi_s)$ following Eq.~3.16 and Eq.~3.17 of Ref.~\cite{Cutler:1997ta}. The polarization angle $\psi_s$ is given by
\ba\label{eq:pol_angle}
\psi_s=\arctan \left(\frac{\hat{L} \cdot \hat{z}-(\hat{L} \cdot \hat{n})(\hat{z} \cdot \hat{n})}{\hat{n} \cdot(\hat{L} \times \hat{z})}\right)\,.
\ea
Here $\hat L$ is the unit vector along the orbital angular momentum of the binary and $\hat z$ is the unit vector along the $z$-axis of the detector. $A_+$ and $A_\times$ are the amplitudes of plus and cross polarization respectively,
\begin{subequations}\label{eq:amp}
\ba
&A_{+}=\frac{2 M_{1} M_{2}}{aD}\left[1+\left(\hat{L}\cdot\hat n\right)^{2}\right]\,, \\
&A_{\times}=-\frac{4 M_{1} M_{2}}{aD} \hat{L}\cdot \hat n\,,
\ea
with $D$ denoting the distance to the Sgr A* EMRI and $a$ is the orbital separation. Various quantities in Eqs.~\eqref{eq:pol_angle}-\eqref{eq:amp} can be expressed in terms of barred coordinates with the help of Eqs.~(3.20)-(3.22) of Ref.~\cite{Cutler:1997ta}.
\end{subequations}

The Doppler phase $\varphi_{D}$ in Eq.~\eqref{eq:waveform} denotes the phase difference of the wavefronts at the detector and at the sun, which is given by
\be\label{eq:Doppler_Phase}
\varphi_{D}(t)=2 \pi f R \sin \bar{\theta}_{S} \cos \left(\bar{\phi}(t)-\bar{\phi}_{S}\right)\,.
\ee
In above expression, $R$ is the distance between sun and the earth, and $\bar \phi(t)$ is associated with location of center-of-mass of the detector:
\be
\bar{\phi}(t)=\bar{\phi}_{0}+2 \pi t / T\,.
\ee
Here $T$ is period of LISA's orbit around sun which is one year, and $\bar{\phi}_{0}$ is a constant denoting the position of the detector at $t=0$ which we will set to zero. Note that the Doppler phase in Eq.~\eqref{eq:Doppler_Phase} is the leading order correction due to the orbital motion of the detector around the sun. Second order Doppler shift to the GW phase is of the order $v|\varphi_D|$, which is a small correction~\cite{Cutler:1997ta} as $v$ is the orbital speed of the detector. Due to the time delay $R\sin\bar\theta$ between a signal at the Barycenter and the detector, there is additional correction to the phase~\cite{Cutler:1997ta}. This is however negligible as it depends on the frequency evolution which is either zero or very small in our analyses. An additional Doppler phase occurs from the rotation of the detector about its axis which is of the order $f$ times the arm length of LISA in geometrized units and such correction is two orders of magnitude smaller than the leading order Doppler correction above. Hence, $\varphi_D$ in Eq.~\eqref{eq:Doppler_Phase} should suffice for our purposes. Finally, we write down  the modulation of GW phase due to orbital precession following~\cite{Apostolatos:1994mx}:
\be\label{eq:phi_p}
\varphi_p(t)=2\int_{0}^{t}\left(\frac{\hat L \cdot \hat{n}}{1-(\hat L \cdot \hat n)^{2}}\right)(\hat L \times \hat n) \cdot \dot{\hat{L}} d t\,,
\ee
where $\dot{\hat{L}}$ is the derivative of $\hat L$ in time.

\subsection{GW phase modulation due to orbital precession of Sgr A* EMRI}\label{subsec:precession_phase}
We now apply Eq.~\eqref{eq:phi_p} for the case of an Sgr A* EMRI. The mass of Sgr A* is $M_1=(4.297\pm0.012)\times10^6 M_{\odot}$~\cite{GRAVITY:2021xju,GRAVITY:2020gka,2018} and we consider a stellar mass compact object with mass $M_2=20 M_{\odot}$. We further assume that the spin of Sgr A* ($S_1$) is much greater than the spin of the stellar mass object ($S_2$), which is reasonable because the mass ratio is already over $10^5$. Consequently, we work in the limit of $M_2\ll M_1$ and $S_2\simeq0$. For such assumptions, $\dot{\hat{L}}$ due to spin precession is given by~\cite{Apostolatos:1994mx}
\be\label{eq:Lhatdot}
\dot{\hat{L}}\simeq\Omega_{p}\,\hat{S}\times \hat{L}\,,
\ee
where the precession frequency $\Omega_p$ takes the form:
\ba
\Omega_P \simeq\frac{2S_1}{a^3}\,.
\ea
Hereafter we will assume an orbital separation of  \red{$a\sim90 M_1$}, in which case the GW radiation-induced phase shift is negligible during the observation period of LISA. One can also find  the dependence of signal-to-noise ratio on the orbital separation in \cite{Gourgoulhon:2019iyu}.
In the well-motivated model where the stellar-mass black holes migrated to the vicinity of the MBH assisted by the MBH accretion disk, the orbital eccentricity damping timescale is much shorter than its migration timescale \cite{Pan2021d}. Therefore we assume that the stellar-mass black hole moves along a circular orbit, although it is straightforward to allow nonzero eccentricity in the Fisher analysis.
The precession angle will be given by
\be
\alpha_p=\alpha_{p,0}+\Omega_P t\,.
\ee
Here $\alpha_{p,0}$ denotes the value of $\alpha$ at time $t=0$. We adopt the convention that $\alpha_p=0$ when $\hat L\cdot \hat{\bar{z}}$ is maximum (which also means $\hat L$, $\hat{\bar{z}}$, and $\hat S$ will be on the same plane), for which $\hat L$ can be written as~\cite{Apostolatos:1994mx}
\begin{align}\label{eq:Lhat}
\hat{L}=\hat{S} \cos \lambda_{p} &+\frac{\left(\hat{\bar{z}}-\hat{S} \cos \bar\theta_{sp}\right)}{\sin \bar\theta_{sp}} \sin \lambda_{p} \cos \alpha_p \nonumber\\
&+\frac{\hat{S} \times \hat{\bar{z}}}{\sin \bar\theta_{sp}} \sin \lambda_{p} \sin \alpha_p\,.
\end{align}
Here $\lambda_p$ is the angle between $\hat L$ and $\hat S$, while $(\bar\theta_{sp},\bar\phi_{sp})$ are spherical polar angles of the vector $\hat S $. The angle $\varphi_p(t)$ now can be evaluated by plugging Eq.~\eqref{eq:Lhat} and Eq.~\eqref{eq:Lhatdot} in Eq.~\eqref{eq:phi_p}.

\subsection{Analysis and results}\label{subsec:data_results}
We now proceed to the discussion of data analysis technique and results pertaining the spin induced orbital precession in the Sgr A* EMRI.
We assume a maximum observation period of $T_{\rm obs}=10$ years with LISA \cite{Audley:2017drz} to explore the science potential of these systems. 
\subsubsection{Data analysis formalism}\label{sec:data}
We implement the Fisher analysis formalism to estimate the statistical error on various parameters in the waveform. Such analysis assumes the noise of the detector to be Gaussian and stationary, which works better in the limit of high SNR. Let us denote the detector's noise as $n(t)$; then the detector output can be expressed in terms of noise and signal as
\be
s(t)=n(t)+h(t)\,.
\ee
We define the inner product between two quantities $A$ and $B$ in the following manner:
\be
(A|B)=4 \Re \int_{0}^{\infty}\frac{\tilde{A}^{*}(f) \tilde{B}(f)}{P_{n}(f)}\,df\,.
\ee
$\tilde A(f)$ is the Fourier transform corresponding to $A$, and the asterisk superscript denotes complex conjugate. $P_n(f)$ is the one-side detector noise spectral density for which we consider the LISA detector noise given in Ref.~\cite{Robson:2018ifk} (we ignore the galactic confusion noise as it is small compared to the instrumental for the frequency range we are working in).
 
The SNR of an event is
\be
\rho=\sqrt{(h|h)}\,,
\ee
while a Fisher matrix element is defined as
\be
\Gamma_{ij}=(\partial_i h|\partial_j h)\,.
\ee
In above equation, we denote a derivative with respect to a binary parameter $\theta^i$ as $\partial_i$. For a Sgr A* EMRI, the orbital frequency can be considered as almost constant during the observation period, so that it is more convenient to use the time-domain waveform which is real-valued. Hence, the Fisher matrix can be simplified in the following way:
\ba
(\partial_i h|\partial_j h)&=&4 \sum_{I, II}\Re \int_{0}^{\infty}\frac{\partial_i\tilde{h}^{*}(f) \partial_j\tilde{h}(f)}{P_{n}(f)}\,df\,,\nonumber\\
&\simeq & \frac{2}{P_n(f)}\sum_{I, II} \int_{0}^{\infty} \partial_i h(t)\partial_j h(t) dt\ ,\nonumber\\
&\simeq & \frac{4}{P_n(f)} \int_{0}^{\infty} \partial_i h(t)\partial_j h(t) dt\ ,
\ea
where we have considered the contribution from two LISA channels $\{I, II\}$, and approximated the Sgr A* EMRI as a monochromatic source with constant frequency $f$. 
In practice, the upper limit of above integral is taken to be the observation period which is 10 years in our case. Let us denote the 1-$\sigma$ error bar on Fisher parameter $\theta^i$ as $\Delta \theta^i=\theta^i_t-\theta^i$ with $\theta^i_t$ denoting the true value of $\theta^i$. If we do not impose any prior information, the root-mean-square of $\Delta \theta^i$ can be obtained as the square root of diagonal elements of the inverse Fisher matrix:
\be
(\Gamma^{-1})^{ab}=\langle\Delta \theta^i\Delta \theta^j\rangle\,.
\ee

The waveform of a precessing Sgr A* EMRI contains \red{nine} parameters as follows:
\be\label{eq:Fisher_Param}
\left(\ln \mathcal A,\bar\theta_{s},\bar\phi_{s},\bar\theta_{sp},\bar\phi_{sp},\alpha_{p,0},\lambda_p,\Omega_p,\mathred{f}\right)\,,
\ee
where $\mathcal A=(M_1 M_2)/(D a)$. Strictly speaking we know the distance $D$ and the sky locations of Sgr A* $(\bar \theta_s,\bar \phi_s)$ from electromagnetic observations, but here we still include $(\bar \theta_s,\bar \phi_s)$ as Fisher variables to address the angular resolution of LISA for such systems.
As the measurement error varies for different system configurations, we use the Monte-Carlo method to sample 100 different sets of underlying parameters of Sgr A* EMRI, assuming uniform sky distribution for the angles and uniform distribution for the spin of Sgr A*. The underlying parameters for $(\bar \theta_s,\bar \phi_s), D, M_1$ are set to be their known values, and \red{$M_2$ is assumed with the same value in the fiducial model ($M_2=20 M_\odot$)}. For each set of the underlying parameters we perform the Fisher analysis to compute the measurement uncertainties for various parameters. The histograms summarizing these Fisher analysis results are presented in Sec.~\ref{sec:results}. 
\subsubsection{Threshold SNR}\label{sec:thre}
In order to address the detectability of Sgr A* EMRIs, we need both the expected SNR of these systems and the SNR detection threshold. The first quantity is partially discussed in \cite{Gourgoulhon:2019iyu} and in Sec.~\ref{sec:results}, and in this section we compute the threshold SNR, following similar strategy described in Ref.~\cite{Moore:2019pke} developed for stellar-mass black hole binaries. Generally speaking, the threshold SNR reflects our tolerance on the false-alarm probability with suitable detection probability. It depends on the number of templates in the template bank against which the observed signal needs to be matched. Let us denote the number of waveforms in the bank by $N_B$, and write the template waveforms as $h_a(t)=\rho \hat h_a(t)$ such that $\langle\hat h|\hat h\rangle=1$, where $a=1, 2,\ldots, N_B$.

The statistics $\sigma_a :=\langle s|h_a \rangle$ between the data $s$ and the templates $h_a$ follows the following probability distribution:
\be
f_1\left(\sigma_a,\rho\right)=\exp\left[-\left(\frac{\sigma_a^{2}+\rho^{2}}{2}\right)\right] \sigma_a I_0(\rho \sigma_a)\,,
\ee
where $I_0$ is the modified Bessel function of the first kind of order $0$. If there is no signal in the data ($s=n$), the 
probability distribution is simply
\be 
f_1(\sigma_a,\rho=0) =  \sigma_a \exp\left[-\left(\frac{\sigma_a^{2}}{2}\right)\right] \ .
\ee 
We can claim a detection if $\sigma_a > \sigma_{\rm thr}$ at least for one template $a$, with $\sigma_{\rm thr}$ denoting a certain threshold determined by a conventional false alarm possibility $P_F$
\be 
P_F(\sigma_{\rm thr}) = \int_{\sigma_{\rm thr}}^{\infty}  f_1\left(\sigma_a, \rho=0\right) d\sigma_a\ ,
\ee 
i.e., $\sigma_{\rm thr}=\sqrt{-2\ln P_F}$.
Following Ref.~\cite{Moore:2019pke}, we choose a representative value of $P_F=10^{-3}/N_B$, which is similar to the threshold $P_F$ in~\cite{LIGOScientific:2016dsl}.  The detection probability as a function of SNR for a given $\sigma_{\rm thr}$ is given by
\be\label{eq:Prob_Dist_SNR}
P_D(\rho)=\int_{\sigma_{\rm thr}}^{\infty}f_1\left(\sigma_a,\rho\right)d\sigma_a\ ,
\ee
and 
the threshold SNR $\rho_{\rm thr}$ is obtained for a given detection probability of $P_D(\rho_{\rm thr})$ which is taken as $0.9$ here. 

We now implement Fisher analyses to estimate the number of templates in the bank $N_B$. 
In our case, the first 3 models parameters 
(the amplitude $\mathcal A$ and the source location $\bar\theta_{s},\bar\phi_{s}$) do not affect $N_B$,
because the amplitude which only shows up as a prefactor of each template, 
therefore can be searched over each template at negligible cost and the source location is well known.
Let us define the following Fisher matrix for the remaining \red{6} model parameters $(\bar\theta_{sp},\bar\phi_{sp},\alpha_{p,0},\lambda_p,\Omega_p,\red{f})$:
\be
\hat \Gamma_{ij}=\left(\partial_i\hat h|\partial_j\hat h\right)\,.
\ee
Let us also define a prior range on a Fisher parameter $\theta^i$ by $\delta \theta^i$, and a modified Fisher matrix $\tilde \Gamma$,
\be
\tilde \Gamma_{ij}={\rm max.} \lb\hat\Gamma_{ij},\frac{\delta_{ij}}{\lb\delta \theta^i\rb^2}\rb\,.
\ee
Then $N_B$ can be approximated as the following integral over the Fisher variables:
\be\label{eq:N_bank}
N_B=\int d^D\theta\sqrt{\mathrm{Det}\, \tilde \Gamma}\,,
\ee
Eq.~\eqref{eq:N_bank} can be evaluated using the Monte-Carlo integration method where the input Fisher parameter $\theta^i$ is distributed over the prior range $\delta \theta^i$~\cite{Cornish:2005hd}:
\be\label{eq:Monte_Carlo_integration}
\int d^D\theta\sqrt{\mathrm{Det}\, \tilde \Gamma} \simeq \frac{V}{N}\sum_{\beta}\sqrt{\mathrm{Det}\,\tilde \Gamma_{\beta}}\,.
\ee
Here $\mathrm{Det}\,\tilde \Gamma_{\beta}$ is the determinant of the of the matrix $\tilde \Gamma$ for sample $\beta$, $N$ is the number of samples, and $V$ is the volume of the search space.

\subsubsection{Results}\label{sec:results}
We first examine the threshold SNR for a detection of GWs from the Sgr A* EMRI. As explained in Sec.~\ref{sec:thre}, for detection purpose, we only need to consider the waveform containing \red{seven} free parameters. In addition, as the cost of searching over amplitude is negligible, we consider only the last \red{six} parameters in Eq.~\eqref{eq:Fisher_Param}. The volume of search space is then $V=\delta\bar\theta_{sp}\,\delta\bar\phi_{sp}\,\delta\alpha_{p,0}\,\delta \lambda_p\,\delta\Omega_p\mathred{\delta f}$. To evaluate Eq.~\eqref{eq:Monte_Carlo_integration}, we apply Monte-Carlo sampling by distributing the direction of spin and angular momentum uniformly over the 2-sphere, $\alpha_{p,0}$ between 0 to $2\pi$, and dimensionless spin parameter between 0.01 to 1. The lower end of spin is set to be nonzero as the Fisher matrix tends to me more singular for small spins, which affects the accuracy in computing the determinant of the Fisher matrix. \red{We distribute the frequency $f$ uniformly between $1.7\times10^{-5}$ Hz to $2\times10^{-5}$ Hz, corresponding to a range of orbital separation $a=85 M_1$ to $a=95 M_1$}. As a consistency check, We implement two independent sets of 100 samples to perform the integration in Eq.~\eqref{eq:Monte_Carlo_integration} and find a consistent threshold SNR $\rho_{\rm thr}\mathred{\approx9.8}$.

In order to estimate the statistical distribution of measurement errors, we perform a separate Monte-Carlo study with the waveform containing all \red{9} parameters in Eq.~\eqref{eq:Fisher_Param}. We distribute $\bar \theta_{sp}$, $\bar \phi_{sp}$, $\alpha_{p,0}$, $\mathred{f}$, and $\lambda_p$ in a similar manner as in the case of threshold SNR computation. However, we distribute the magnitude of dimensionless spin parameter from 0 to 1 uniformly as the calculation of determinant is not needed here. For each set of Monte-Carlo samples we compute the Fisher matrix, and collect all the data in histograms.

\begin{figure}
\includegraphics[width=8.5cm]{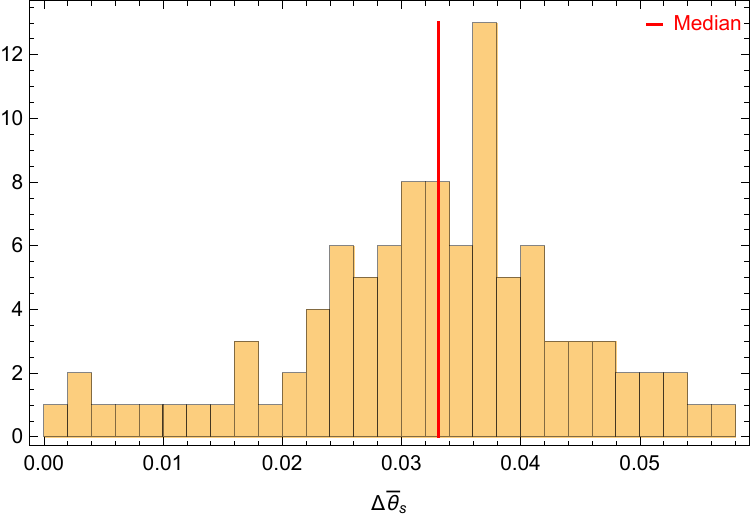}
\caption{Histogram distribution of 1-$\sigma$ errors on the polar angle of location of Sgr A* EMRI ($\Delta \bar \theta_s$) with respect to the heliocentric coordinates, computed with Fisher analyses with Monte-Carlo simulations. Vertical red line is the median of distribution which is about 0.033 radians.}
\label{fig:ThetaS}
\end{figure}

\begin{figure}
\includegraphics[width=8.5cm]{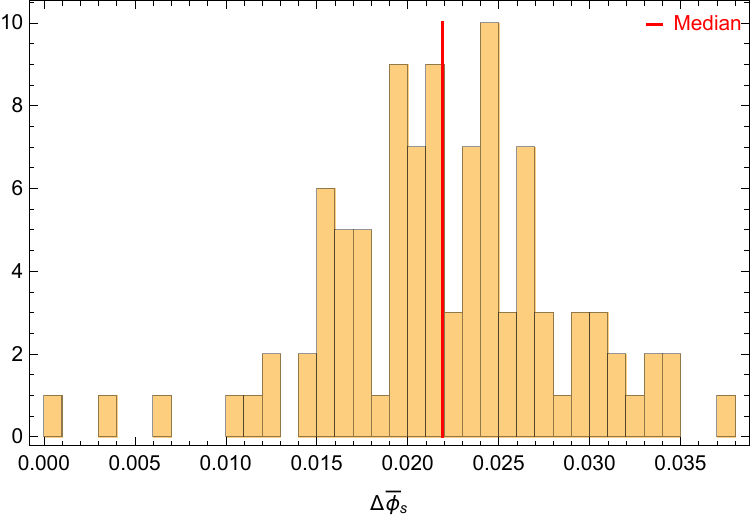}
\caption{Histogram distribution of 1-$\sigma$ errors on the azimuthal angle of location of Sgr A* EMRI ($\Delta \bar \phi_s$) in the heliocentric coordinates, computed with Fisher analyses with Monte-Carlo simulations. Vertical red line is the median of distribution which is about 0.022 radians.}
\label{fig:PhiS}
\end{figure}

\begin{figure}
\includegraphics[width=8.5cm]{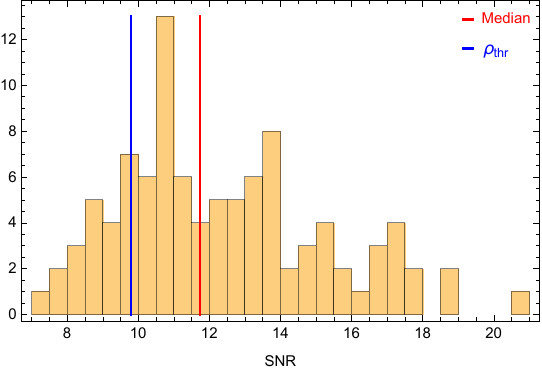}
\caption{Histogram distribution of SNRs obtained from Fisher analyses with Monte-Carlo simulations. The blue vertical line shows the threshold SNR for detecting a signal: $\rho_{thr}\simeq \mathred{9.8}$. The red vertical line shows the median SNR which is \red{11.7}. About $\mathred{80\%}$ of the samples satisfy $\rho \geq\rho_{thr}$.}
\label{fig:SNR}
\end{figure}

\begin{figure}
\includegraphics[width=8.5cm]{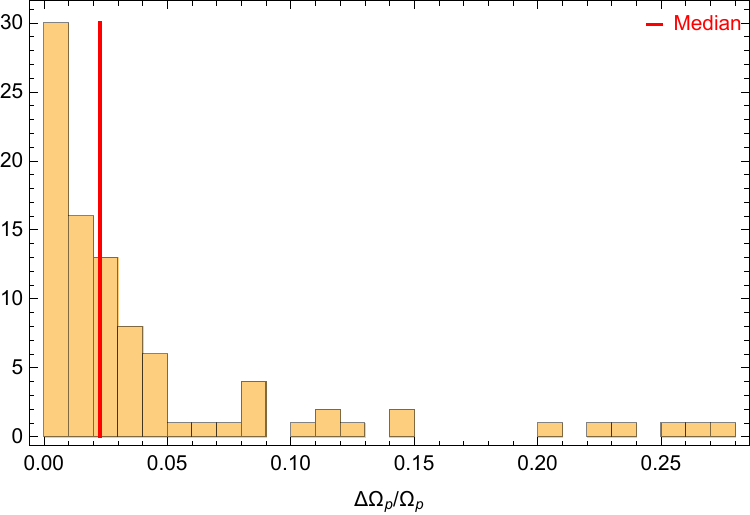}
\caption{Histogram distribution of relative errors in measurement of precession orbital frequency $\Omega_p$ obtained from Fisher analyses with Monte-Carlo simulations. The vertical red line denotes the median value of $\Delta \Omega_p/\Omega_p$ which is approximately 0.02.}
\label{fig:OmegaP}
\end{figure}

\begin{figure}
\includegraphics[width=8.5cm]{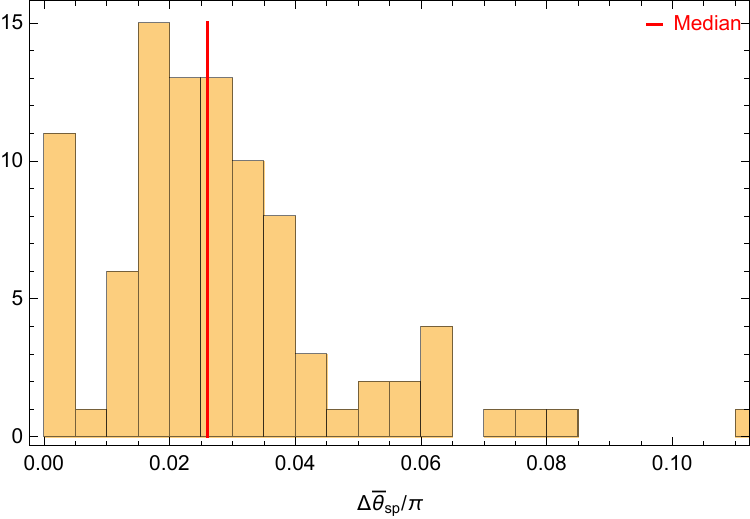}
\caption{Histogram distribution of 1-$\sigma$ error bars on the polar angle corresponding to the direction of spin ($\Delta \bar \theta_{sp}$), normalized by $\pi$. Vertical red line shows the median of the distribution which is approximately \red{0.026}.}
\label{fig:ThetaSp}
\end{figure}

\begin{figure}
\includegraphics[width=8.5cm]{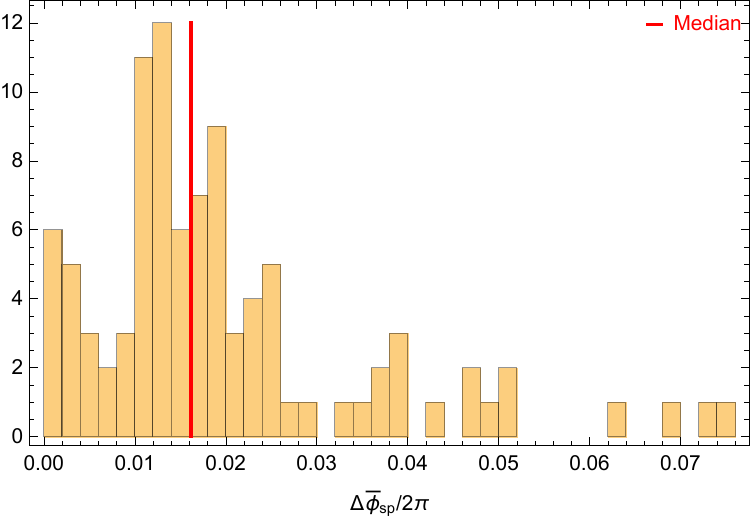}
\caption{Histogram distribution of 1-$\sigma$ error bars on the azimuthal angle corresponding to the direction of spin ($\Delta \bar \phi_{sp}$), normalized by $2\pi$. Vertical red line shows the median of the distribution which is approximately \red{0.016}.}
\label{fig:PhiSp}
\end{figure}

\begin{figure}
\includegraphics[width=8.5cm]{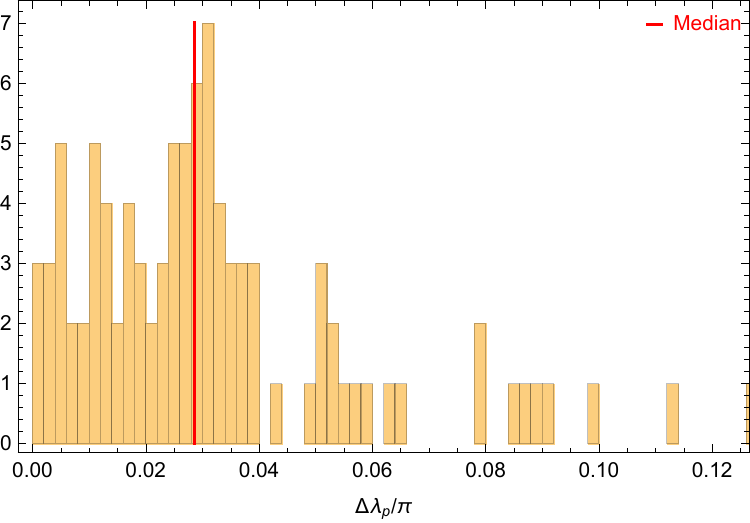}
\caption{Histogram distribution of 1-$\sigma$ errors in the measurement of opening angle ($\Delta\lambda_p$) normalized by $\pi$ obtained from Fisher analyses with Monte-Carlo simulations. The vertical red line shows the median value of $\Delta\lambda_p/\pi$ which is approximately \red{0.03}.}
\label{fig:LambdaP}
\end{figure}

\begin{figure}
\includegraphics[width=8.5cm]{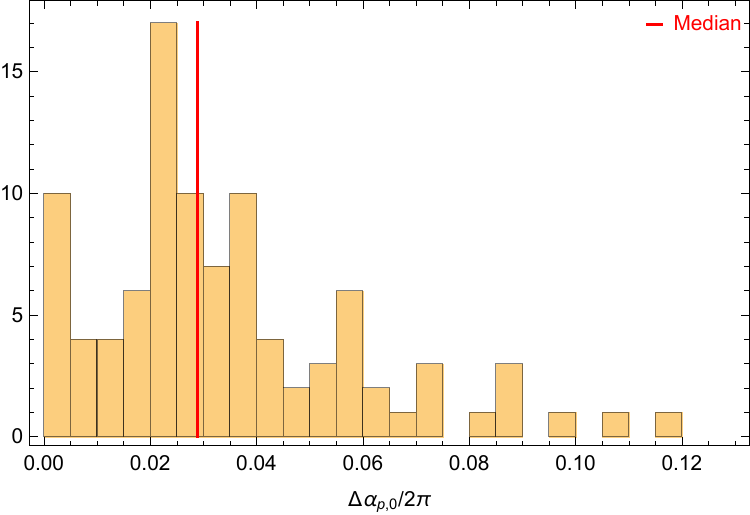}
\caption{Histogram distribution of 1-$\sigma$ error bars in the measurement of initial precession angle ($\Delta\alpha_{p,0}$) normalized by $2\pi$. Vertical red line shows the median value of the distribution which is approximately 0.03.}
\label{fig:Alpha0}
\end{figure}

\red{For the $M_2$ and range of $a$ assumed in this fiducial model, we find 80\% of the samples produce SNR greater than the threshold SNR $\rho_{thr}$}, while the median SNR is about 12 (see Fig.~\ref{fig:SNR}). Note that the SNR linearly scales as $M_2$ if it is different from $20 M_\odot$, and its dependence on $a$ is discussed in \cite{Gourgoulhon:2019iyu}. We also present a histogram distribution of the relative errors in measuring $\Omega_p$ in Fig.~\ref{fig:OmegaP}, which shows that we can constrain $\Omega_p$ (and hence the magnitude of spin) with a median relative error of 2\%. The direction of spin can also be measured to percent level accuracy, as shown in Fig.~\ref{fig:ThetaSp} and \ref{fig:PhiSp}. This level of accuracy is more than an order of magnitude better than the spin measurement accuracy of radio interferometry, which has to account for all the modelling uncertainties of the accretion flow. In addition, Fig.~\ref{fig:ThetaS}  and Fig.~\ref{fig:PhiS} show that the LISA resolution uncertainties for the sky angles are on the level of $\Delta \bar \theta_{s} \sim 0.033, \Delta  \bar{\phi_{s}} \sim 0.022$ radian, or $\Delta \Omega/4\pi \sim 6\times 10^{-5}$. This means that the chance of having a double white dwarf source having similar signal strength (less than parsec scale distance) and sky locations is negligible. The expected measurement error of the opening angle $\lambda_p$ and initial precession angle $\alpha_p$ are shown in Fig.~~\ref{fig:LambdaP}, and~\ref{fig:Alpha0}. \red{On the other hand, the frequency can be measured very precisely with an error bar of the order $10^{-10}$ Hz}.


\section{Probing dark matter}\label{sec:dark_matter}
Massive black holes like Sgr A* can harbor dark matter clouds around them, and one way to probe the nature of such clouds is through GW observations. When a compact object travels through the cloud, a dark matter overdensity may form and create a friction force. Such friction force changes the frequency evolution of the binary. The cloud's gravitational potential  also applies a torque on the compact object, giving rise to an orbital precession. This section will focus on detecting such effects with Sgr A* EMRI, with  both cold dark matter and axionic dark matter scenarios considered.
\subsection{Cold dark matter}\label{sec:dark_matter_cold}
Dark matter may accumulate around a massive black hole like Sgr A*, forming overdensities or spikes ~\cite{Gondolo:1999ef,Ullio:2001fb}.
The change to GW phase due to the dynamical friction may be used to probe the dark matter profile around the central massive object, which has been explored in the case of intermediate-mass-ratio inspirals (IMRIs) ~\cite{Kavanagh:2020cfn,Coogan:2021uqv,Eda:2014kra}. In the analysis below we comment on its application on EMRIs, especially the Sgr A* EMRIs.

Let us denote the frequency evolution as $\dot f$, which is small and can be treated perturbatively with respect to the frequency of GWs at the beginning of observation period: $f_0$. Using the energy-balance law, $\dot f$ can be found as~\cite{Kavanagh:2020cfn,Coogan:2021uqv}
\be\label{eq:coldDM_freq_ev}
\dot f=\frac{96 \pi ^{8/3} f_0^{11/3}M_1M_2}{5M^{1/3}}+\frac{12 M_2\xi(v)\rho_0^{DM} \ln (\Lambda )}{M_1}\,.
\ee
The first term on the right side of above equation is the frequency evolution due to the emission of GWs, while the second term is the contribution coming from dynamical friction. Here $\Lambda$ is defined as $\sqrt{M_1/M_2}$ and $\xi (v)$ is the fraction of dark matter moving slower than the orbital speed $v$. The dark matter density at the moving object's location  at time $t=0$ is $\rho_0^{DM}$:
\be\label{eq:coldDM_profile}
\rho^0_{DM}=\rho_{sp}\left(\frac{a_{sp}}{a_0}\right)^{\gamma_{sp}}\,,
\ee
with $a_0$ denoting the orbital separation at $t=0$. The size of the spike is $a_{sp}$ and $\rho_{sp}$ is the density of dark matter at distance $r=a_{sp}$. The parameters ($a_{sp},\rho_{sp},\gamma_{sp}$) depend on the initial dark matter halo where the supermassive black hole was formed. If we consider the initial profile to be an NFW profile, $\gamma_{sp}$ is usually taken  as $7/3$. Using Eq.~\eqref{eq:coldDM_profile} to Eq.~\eqref{eq:coldDM_freq_ev}, we find that the contribution to $\dot f$ from the dynamical friction scales as $f_0^{14/9}$. As $f_0$  for Sgr A* EMRIs is smaller than that of ordinary EMRIs detected by LISA by two orders of magnitude, the dark matter induced modification to $\dot f$ and GW phase (which scales as $\dot{f} T^2_{\rm obs}$) is supposed to be two orders of magnitude smaller as well. We expect such dark matter induced corrections to be more significant for ordinary EMRIs with smaller separation, higher orbital frequency and thus  better detection prospects.

For a quantitative assessment of the modification to GW frequency evolution due to dynamical friction, let us first derive the parameters $(a_{sp},\rho_{sp})$ for the cloud profile around Sgr A*. For the initial NFW profile, DM density $\rho$ at a distance $r$ from Sgr A* is~\cite{Navarro:1995iw}
\be\label{eq:NFW}
\rho_{\mathrm{NFW}}(r)=\frac{\rho_{s}}{\left(r / a_{s}\right)\left(1+r / a_{s}\right)^{2}}\,,
\ee
where $\rho_s$ and $a_s$ are characteristic density and radius respectively. $a_s$ is usually defined to be the radius within which the average DM density becomes 200 times the critical density of the universe $\rho_c$. Then $\rho_s$ is related to $\rho_c$ according to
\be\label{eq:rho_s}
\frac{\rho_{s}}{\rho_c}=\frac{200}{3} \frac{c_{200}^{3}}{\mathcal F\left(c_{200}\right)}\,,
\ee
where $\mathcal F(x)=\ln (1+x)-x/(1+x)$~\cite{Dutton:2014xda}. $c=r/a_s$ is the concentration parameter, and DM mass within $r=a_s$ is approximated by
\be
M_{200}=4\pi \rho_s a_s^3 \mathcal F(c_{200})\,.
\ee
We consider the low red shift approximation for parameter $c_{200}$ which is~\cite{Dutton:2014xda}
\be
\log _{10} c_{200}=0.905-0.101 \log _{10}\left[\frac{M_{200} h}{10^{12}M_{\odot}}\right]\,,
\ee
where $h$ is the dimensionless Hubble constant and $M_{200}$ is related to Sgr A* mass $M_1$ as~\cite{Ferrarese:2002ct}
\be\label{eq:M_200}
\frac{M_1}{10^{7} M_{\odot}}\approx \left(\frac{M_{200}}{10^{12} M_{\odot}}\right)^{1.65}
\ee
Using Eqs.~\eqref{eq:rho_s}-\eqref{eq:M_200},we achieve $a_s=19.6$ kpc and $\rho_s=3 \times 10^{-28}\mathrm{g/cm^3}$.

Next, following~\cite{Merritt:2003qk} we take the spike size $a_{sp}$ to be $0.2r_h$ where $r_h$ is the gravitational influence radius of Sgr A* which is 1.67pc. Then matching Eq.~\eqref{eq:NFW} to spike profile $\rho_{DM}=\rho_{sp}(a_{sp}/r)^{7/3}$ at $r=a_{sp}$ we obtain $\rho_{sp}\simeq 262 \mathrm{M_{\odot}/pc^3}$. Finally, Let us define the second term in Eq.~\eqref{eq:coldDM_freq_ev} as
\be
\dot f_{DF}=\frac{12 M_2\xi(v)\rho_0^{DM} \ln (\Lambda )}{M_1}
\ee
Plugging $\rho_{sp}$ and $a_{sp}$ in above equation, and considering static dark matter case ($\xi=1$) for simplicity, we obtain $\dot f_{DF}=4\times 10^{-21}s^{-2}$. This is more than two orders of magnitude smaller compared to the frequency evolution due to GW emission which we find as $1.3\times 10^{-18}s^{-2}$. We present a comparison of $\dot f_{DF}$ between the cold DM case and Axion DM case in the next subsection.
\subsection{Axion dark matter}\label{sec:dark_matter_axion}
Axion or axion-like particles are well-motivated from string theory~\cite{Marsh:2015xka,Arvanitaki:2009fg} and standard model extension~\cite{Peccei:1977hh,Weinberg:1977ma,Dine:1982ah,Abbott:1982af} as dark matter candidates~\cite{Press:1989id,Peebles:2000yy,Amendola:2005ad,Hui:2016ltb,Preskill:1982cy}. Axion fields close to spinning BH can extract angular momentum from the BH, leading to superradiance. In this subsection, we consider the effect of a bosonic cloud around Sgr A* on the orbit of the stellar mass BH. Axion dark matter density profile can achieve extremely high concentration near the central SMBHB, as compared with the cold dark matter scenario. Additional conservative dynamical effect may become significant because of the high mass density. For example, the Newtonian potential of the cloud can exert a torque on the orbiting object and induce orbital precession. We first study such precession effect in case of an Sgr A* EMRI, and then we analyze the change in GW frequency evolution induced by the dynamical friction of axion dark matter.
\subsubsection{Newtonian potential of the cloud}
To study the nonrelativistic description of the superradiant cloud, we will work in spherical coordinate systems $(r,\theta,\phi)$ and Cartesian coordinate $(x,y,z)$ with Sgr A* located at the origin. We consider a free complex scalar field, in which case the cloud is axisymmetric around the direction of spin of Sgr A*~\cite{Zhang:2019eid,Baumann:2018vus}. In the nonrelativistic limit, the axion field can be expressed as
\be\label{eq:WaveFunction}
\Psi(r,t)=\mathcal B e^{-i \omega t} R_{nl}(r) Y_{\ell m}(\theta, \phi)\,.
\ee
$(n,l,m)$ label the stationary eigenmodes of the field and $\omega$ is the eigenfrequency. The radial function $R_{nl}$ takes the form of that of hydrogen atom, which we can conveniently write as
\ba
R_{n \ell}(\tilde x)=&&\left[\left(\frac{2\alpha \mu}{n}\right)^{3}\frac{(n-\ell-1) !}{2 n(n+l) !}\right]^{1 / 2}\times\nonumber\\
&&e^{-\frac{\tilde x}{n}}\left(\frac{2 \tilde x}{n}\right)^{\ell}  L_{n-\ell-1}^{2 \ell+1}\left[\frac{2 \tilde x}{n}\right]\,.
\ea
Here we introduce the parameter $\tilde x=\alpha^2 r/M_1$ with $\alpha=M_1 \mu$ and $\mu$ is the mass of the scalar field. $L_{n-\ell-1}^{2 \ell+1}\left[\frac{2\tilde x}{n}\right]$ is the generalized Laguerre polynomial of degree $(n-l-1)$~\footnote{We implement the convention where $L_n^a(x)$ satisfies the differential equation $xy''+(a+1-x)y'+ny=0$.}. We choose the constant $\mathcal B$ in Eq.~\eqref{eq:WaveFunction} such that the total mass of the cloud $M_c$ is normalized to $\alpha M_1$, namely,
\be\label{eq:normalization}
\int d^{3} x \rho=M_{c}\sim \alpha M_1\,,\qquad\rho=\mu \Psi^{*} \Psi\,.
\ee
Such an approximation is reasonable when $\alpha\ll 1$ and the initial BH spin is close to the maximal spin~\cite{Baumann:2018vus}. Using Eq.~\eqref{eq:normalization} we then find $\mathcal B=M_1$.

In the next step we derive the gravitational potential generated by the bosonic cloud. Since the potential satisfies the Poisson equation, we decompose it in terms of spherical harmonics in the following manner~\cite{2014grav.bookpoissonwill,Ferreira:2017pth}:
\ba\label{eq:potential}
\Phi_{c}=\sum_{\ell m} \frac{4 \pi }{2 \ell+1}&&\left[q_{\ell m}(t, r) \frac{Y_{\ell m}(\theta, \phi)}{r^{\ell+1}}\right.\nonumber\\
&&\left.+p_{\ell m}(t, r) r^{\ell} Y_{\ell m}(\theta, \phi)\right],
\ea
with
\begin{subequations}\label{eq:qp}
\ba
q_{\ell m}(t, r) &=\int_{0}^{r} s^{\ell} \rho_{\ell m}(s) s^{2} d s\,, \\
p_{\ell m}(t, r) &=\int_{r}^{\infty} \frac{\rho_{\ell m}(s)}{s^{\ell+1}} s^{2} d s\,.
\ea
\end{subequations}
We further define the harmonic components of the density $\rho$ as
\be\label{eq:rholm}
\rho_{\ell m}(t, r)=\int \rho(r, \theta, \phi) Y_{\ell m}^{*}(\theta, \phi) d \Omega\,.
\ee

For simplicity we restrict ourselves to the fastest growing mode, which corresponds to $n=2$, $l=1$, and $m=1$. For this particular mode, using Eq.~\eqref{eq:WaveFunction} and Eq.~\eqref{eq:normalization} to Eq.~\eqref{eq:rholm}, non-zero contributions to cloud density $\rho$ come from
\begin{subequations}\label{eq:rholmfor221}
\ba
\rho_{00}&=&\frac{M_1^2\tilde x^2 e^{-\tilde x} \alpha^3\mu^4}{48 \sqrt{\pi}}\,,\\
\rho_{20}&=&-\frac{M_1^{2}\tilde x^2 e^{-\tilde x} \alpha^3\mu^4}{48 \sqrt{5\pi}}\,.
\ea
\end{subequations}
Finally, using Eq.~\eqref{eq:rholmfor221} and Eq.~\eqref{eq:qp} to Eq.~\eqref{eq:potential}, we obtain the gravitational potential of the cloud as
\be\label{eq:potential2}
\Phi_c=P_1(\tilde x)+P_2(\tilde x)\cos{2\theta}\,,
\ee
with
\ba
P_1=\frac{\alpha^{3} e^{-\tilde x}}{32\tilde x^{3}}(48-48 e^{\tilde x}&+&48 \tilde x-8\tilde x^{2}+32 e^{\tilde x} \tilde x^{2}\nonumber\\
&-&16 \tilde x^{3}-6\tilde x^{4}-\tilde x^{5})\,,
\ea
and
\ba
P_2=\frac{\alpha^{3}e^{-\tilde x}}{32\tilde x^{3}}(144-144 e^{\tilde x}&+&144 \tilde x+72 \tilde x^{2}\nonumber\\
&+&24 \tilde x^{3}+6 \tilde x^{4}+ \tilde x^{5})\,.
\ea
\subsubsection{Axion dark matter induced precession}
It is convenient to express the components of the torque applied on Sgr A* EMRI orbit using the  Cartesian coordinates $(x,y,z)$. Implementing the Newtonian potential in Eq.~\eqref{eq:potential2}, the torque on object $M_2$, is
\ba\label{eq:torque}
\vec\tau&=&-M_2\vec r \times \vec \nabla \Phi_c\nonumber\\
&=&-4 M_2 P_2(\tilde x)\{\sin\theta \cos\theta \sin\phi,-\sin\theta\cos\theta\cos\phi,0\}\,.\nonumber\\
\ea
Such a torque will cause the orbit of $M_2$ to precess around the direction of spin of Sgr A*. However, since the precession frequency is expected to be much smaller than the orbital frequency, we are interested in the torque averaged over one period. In this case the point mass can be replaced by a constant density ring of the same radius to receive the gravitational torque. Let us choose the  z-axis to be along the direction of spin  and the x-axis to be along the ascending node of the binary. We also assume at $t=0$ object $M_2$ is located along x-axis.  The position of $M_2$ can be expressed  as
\ba
x=a \cos\Omega t,\qquad y=a \cos \lambda_p\sin\Omega t,\qquad z=a \sin \lambda_p\sin\Omega t\,.\nonumber\\
\ea
Here $\Omega$ is the orbital angular frequency and $\lambda_p$ is the angle between z-axis and the orbital angular momentum. Applying above equation to Eq.~\eqref{eq:torque} and averaging over one orbital period we obtain
\be\label{eq:avg_torque}
\langle\vec\tau\rangle=\left\{-P_2(\tilde x(a)) M_2\sin2\lambda_p,0,0\right\}\,.
\ee
Next, we can express the precession frequency as a small change of precession angle $\Delta \Theta$ over small time $\Delta t$ as
\be
\Omega_p=\frac{\Delta\Theta}{\Delta t}\simeq \frac{\tan{\Delta\Theta}}{\Delta t}=\frac{\Delta\vec{L}}{L\sin{\lambda_p}\Delta t}=\frac{|\langle\vec \tau\rangle|}{L\sin \lambda_p}\,.
\ee
Using Eq.~\eqref{eq:avg_torque} to above equation leads to a precession frequency of
\be\label{eq:Omega_p}
\Omega_p=\frac{4\Omega^{1/3}P_2\cos\lambda_p}{M_1^{2/3}}\,,
\ee
which depends on $\alpha$ through $P_2$. We need to choose a suitable range for $\alpha$, which we discuss next.

To allow a detectable effect for Sgr A* EMRI, the growth period of the superradiant cloud needs to be much smaller than the Hubble time. In the limit of $\alpha\ll 1$, the growth period of the cloud is given by the Detweiler's approximation~\cite{Baumann:2018vus,Detweiler:1980uk} :
\be
\tau_g \simeq 24(a / M_1)^{-1} \alpha^{-(4 \ell+5)} M_1\,.
\ee
If we set $\tau_g$ to be $\mathcal O (10^8)$ years, above equation suggests $\alpha\simeq 0.04$, which we will consider as a lower limit on $\alpha$. On the other hand, the cloud may deplete due to GW radiation if the field is real-valued. Notice that in this case the calculation for precession rate still applies, as the cloud rotates faster than the precession rate, and a rotation-averaged density profile for real-valued field is the same as that for complex-valued field.  Taking into account the GW radiation, the lifetime of the cloud can be approximated by the following expression which is valid for $\alpha<0.1$~\cite{Baumann:2018vus}:
\be
\tau_c\simeq 10^{9} \text { years }\left(\frac{M}{10^{5} M_{\odot}}\right)\left(\frac{0.1}{\alpha}\right)^{15}\,.
\ee
If we choose an $\alpha$ of 0.1, which is the limit of validity of above expression, the cloud survives about 1 billion years for the Sgr A* EMRI. Hence it is reasonable to assume 0.1 as the upper bound on $\alpha$. A range of $\alpha \in [0.04,0.1]$ corresponds to the mass of the scalar field $\mu \in [13,32.6]\times 10^{-19}$ eV.

\begin{figure}[htb]
\includegraphics[width=8.5cm]{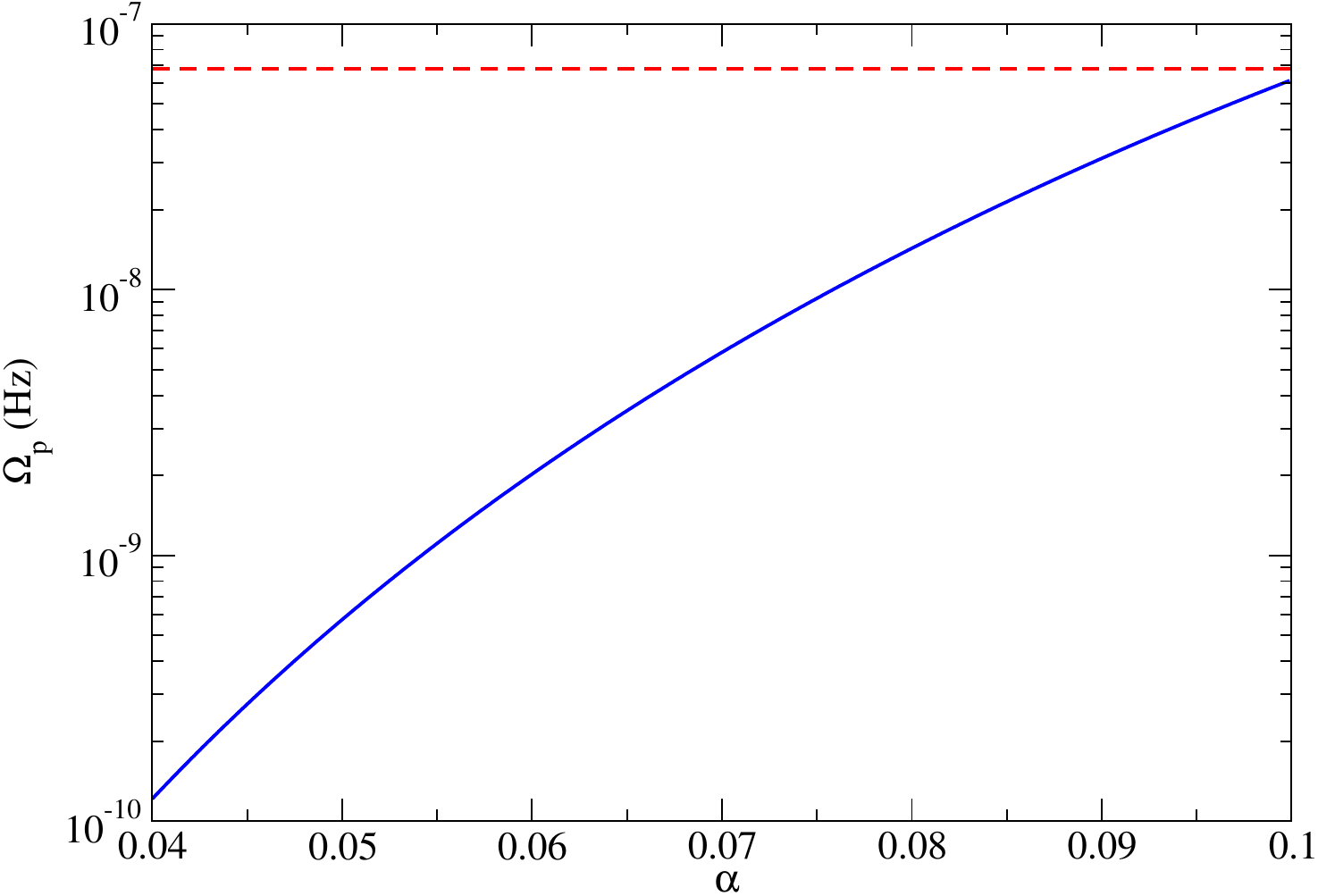}
\caption{Axion dark matter induced precession orbital frequency $\Omega_p$ as a function of $\alpha= M_1\mu$ according to Eq.~\eqref{eq:Omega_p}, which is denoted by the blue curve. Here $M_1$ is the mass of Sgr A* and $\mu$ is the mass of the complex scalar field which describes the axion dark matter. The angle between spin and orbital angular momentum vector is taken to be $\pi/4$. The red horizontal dotted line shows the orbital frequency of spin induced precession with same opening angle and $S=0.5 M_1^2$.}
\label{fig:axiondom}
\end{figure}

In Fig.~\ref{fig:axiondom} we present $\Omega_p$ as a function of $\alpha$ for $\lambda_p=\pi/4$, and also compare it with spin induced precession with same opening angle and a spin of $S=0.5 M_1^2$. We find that the dark matter induced precession frequency is overall smaller than that of spin-induced precession, although they are almost comparable near $\alpha=0.1$. Nevertheless, since in both cases the precession happens around the direction of spin, the modification to GW phase is similar and in principle it is not possible to distinguish the two effects.
\subsubsection{Modification to frequency evolution due to dynamical friction}
In addition to the orbital precession caused by the Newtonian potential of the axion cloud, a  dynamical friction force is also applied on object $M_2$, which can be expressed as~\cite{Hui:2016ltb,1943ApJ....97..255C}
\be\label{eq:friction_force}
F=\frac{1}{v^2}4\pi M_2^2\rho C_{\Lambda}\,.
\ee
In case of Sgr A* EMRI where the orbit is circular, the velocity $v$ of $M_2$ relative to the wave function can be approximated as the orbital velocity, and $C_{\Lambda}$ has a simplified form~\cite{Zhang:2019eid}
\be\label{eq:Clambda}
C_{\Lambda}(k r)\simeq \int_0^{2 k r}\frac{1}{s}(1-\cos s)ds +\frac{\sin 2 k r}{2 k r}-1\,,
\ee
where $k=\mu v$ with $\mu$ representing the mass of the scalar field.
We again restrict ourselves to the $n=2$, $l=1$, and $m=1$ mode and use Eq.~\eqref{eq:Clambda} and $\rho=\Psi^*\Psi$ to Eq.~\eqref{eq:friction_force} to obtain the rate of change of energy loss due to dynamical friction as $\dot E _{DF}=Fv$. Similar to \ref{sec:dark_matter_cold}, Energy loss due to dynamical friction and GW emission in this case can be taken as small and constant, which enters the GW phase as $\delta\varphi=\dot f \pi t^2$. The waveform we consider in this case is
\ba
h(t)=\frac{\sqrt{3}}{2} A(t) \cos \left[2\pi f_0 t+\dot f \pi t^2+\varphi_{p}(t)+\varphi_{D}(t)\right]\,,\nonumber\\
\ea
where $\varphi_p (t)$ is the polarization phase and $\varphi_D(t)$ is the Doppler phase as in section~\ref{subsec:waveform}, and
\be
\dot f\simeq\frac{96 \pi ^{8/3} f_0^{11/3}M_1M_2}{5M^{1/3}}+\frac{3f_0^{1/3}M^{1/3}F_0v_0}{\pi^{2/3}M_1 M_2}\,.
\ee
Quantities with subscript ``0''  are evaluated for a circular binary with $a=90M$. The second term on the right side is the contribution to frequency evolution from dynamical friction: $\dot f_{DF}$.

\begin{figure}[htb]
\includegraphics[width=8.5cm]{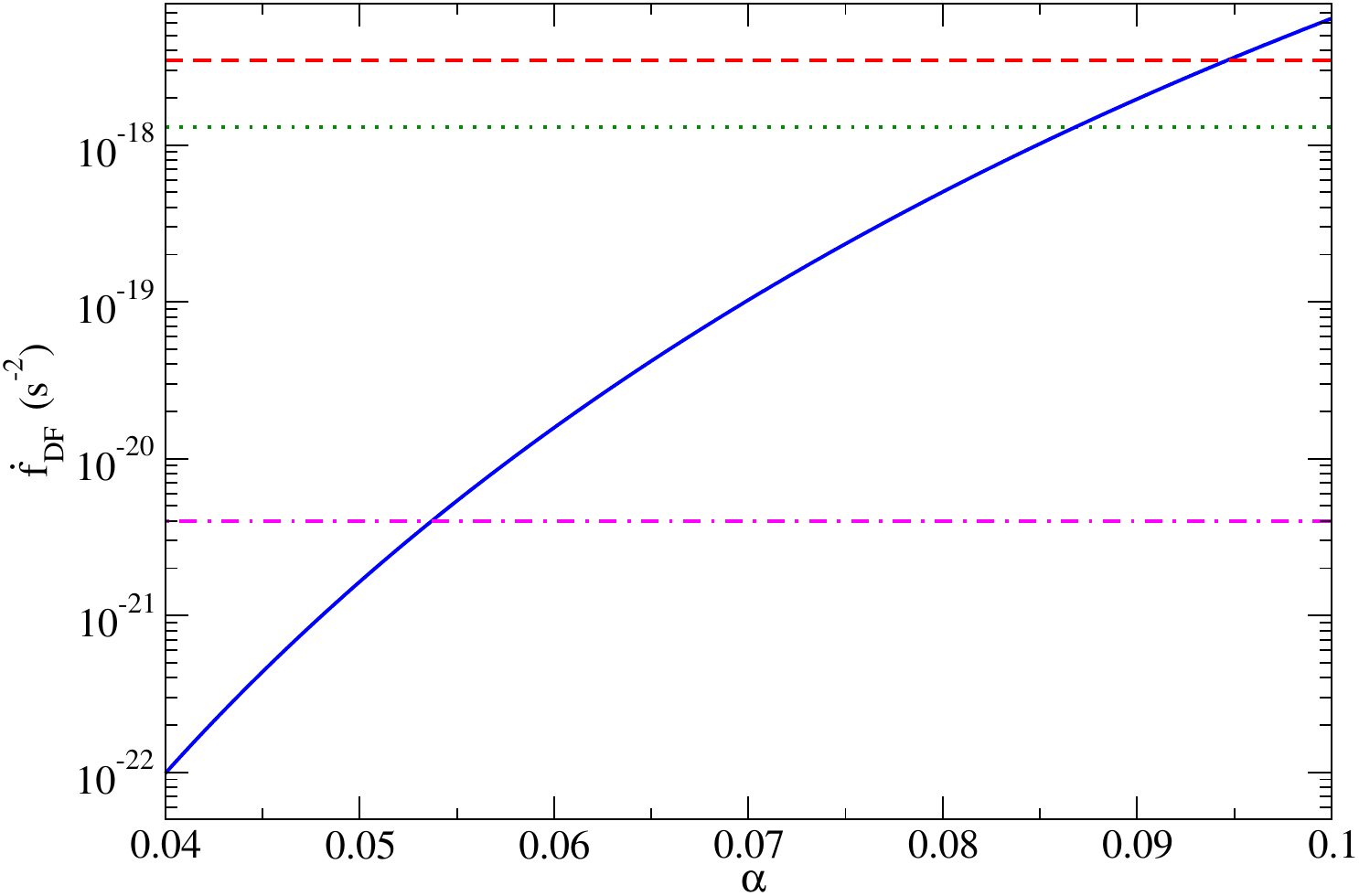}
\caption{Dynamical friction induced frequency evolution ($\dot f_{DF}$) as a function of $\alpha$ (shown by blue solid line) for a Sgr A* EMRI in equatorial plane $\theta=\pi/2$. The horizontal dashed red line shows the 1-$\sigma$ upper bound on $\dot f_{DF}$ achieved from Fisher analyses, which roughly corresponds to $\Delta \dot f_{DF}=\mathred{3.47\times 10^{-18}}\,\mathrm{s}^{-2}$. The green dotted line indicates the frequency evolution due to emission of GWs ($1.3\times10^{-18}s^{-2}$), while the magenta dash-dotted line represents $\dot f_{DF}$ in the case of a cold dark matter spike around Sgr A* ($4\times10^{-21}s^{-2}$) discussed in Sec.~\ref{sec:dark_matter_cold}.
}
\label{fig:freq_ev}
\end{figure}

In Fig.~\ref{fig:freq_ev} we present $\dot f_{DF}$ as a function of $\alpha$ for a Sgr A* EMRI in the equatorial plane $\theta=\pi/2$, as well as the 1-$\sigma$ upper bound on $\dot f_{DF}$ obtained from Fisher analyses. The Fisher analysis is performed with the parameters $\ln \mathcal A$, $\bar \theta_s$, $\bar \phi_s$, $\bar \theta_L$, $\bar\phi_L$, $\mathred{f_0}$, and $\dot f_{DF}$. Here ($\bar\theta_L,\bar\phi_L$) is the direction of orbital angular momentum of Sgr A* EMRI, for which we consider ($\pi/4,3\pi/4$) as a fiducial value. The error of $\dot f_{DF}$ is not sensitive to the fiducial value of $\dot f_{DF}$. We find the 1-$\sigma$ upper bound on $\dot f_{DF}$ is approximately $\mathred{3.47\times 10^{-18}}\,\mathrm{s}^{-2}$, which roughly corresponds to $\mathred{\alpha=0.095}$, which means an effect to the waveform phase coming from  $\mathred{\alpha \in [0.095,0.1]}$ or $\mathred{\mu \in [3.1,3.3]\times 10^{-18}}$ eV can exceed the statistical uncertainty. The chance that axion mass lies within such narrow range is small. Fig.~\ref{fig:freq_ev} also shows that the dynamical friction due to cold DM spike around Sgr A* can produce $\dot f_{DF}$ which is comparable to the Axion DM case only when $\alpha\leq 0.054$. Note that this is an overestimate, as we assumed static cold dark ($\xi=1$) while in reality $\xi$ is expected to be smaller~\cite{Coogan:2021uqv,Kavanagh:2020cfn}. On the other hand, $\dot f_{DF}$ in Axion DM case can be greater than the GW radiation induced $\dot f$ if $\alpha>0.087$.
\section{Discussion}\label{sec:discussion}
\subsection{Multi-body effect }
We now explore the case where the Sgr A* EMRI experiences the Kozai-Lidov effect due to the presence of a tertiary stellar-mass object, and determine whether or not such effects would be detectable. Let us denote the mass of the third body by $M_3$, which orbits around the center of mass of the Sgr A* EMRI consisting of masses $M_1$ and $M_2$. We assume $a_{out}\gg a_{in}$ where $a_{in}$ and $a_{out}$ are semimajor axes of the inner and outer binary respectively. We also assume the eccentricities of inner and outer binaries  are $e_{in}$ and $e_{out}$ respectively. The outer perturber $M_3$ induce time-dependent  evolution of the orbital elements of the inner binary, such as the eccentricity, inclination, argument of periapsis, and longitude of ascending nodes.

The orbital evolution of the Sgr A* EMRI in such a hierarchical triple system happens in a typical rate  $t_{LK}^{-1}$, where $t_{LK}$ is Kozai-Lidov time scale~\cite{2014,Su:2020vda}:
\be
t_{LK}^{-1} =\sqrt{\frac{M_1+M_2}{a_{in}^3}}\left(\frac{M_{3}}{M_1+M_2}\right)\left(\frac{a_{in}}{a_{out}\sqrt{1-e_{out}^2}}\right)^{3}\,.
\ee
In this case, $M_3/(M_1+M_2) \sim \mathcal{O}(10^{-5})$, together with the fact that $\sqrt{(M_1+M_2)/a_{in}^3} \sim \mathcal O (10^{-5})$ Hz, suggests that $t_{LK}^{-1} \sim \mathcal{O}(10^3)$ year, i.e., the Kozai-Lidov time scale is much longer than our observation period.
Therefore the multibody effect is negligible for the Sgr A* EMRIs considered here.
\subsection{Conclusion}
We have investigated several possible applications of Sgr A* EMRIs if they exist within $10^2$M distance from Sgr A*. The detection threshold for these systems is significantly smaller than that of ordinary EMRIs, thanks to the known location of the source and much smaller waveform parameter space. The primary application of an Sgr A* EMRI is to measure the spin of Sgr A*, which likely has the best accuracy compared to alternative observation methods, such as using EHT. We have investigated the case of a cold dark matter cloud around Sgr A* and have found that EMRIs that have higher orbital frequency would be more efficient to probe such a scenario. On the other hand, the presence of an axion cloud, if the axion mass falls into the range with $\alpha \in (0.04-0.1)$, may generate an additional precession effect for the EMRI orbit, which is degenerate with the spin induced precession. However, the degeneracy effect is only significant  in a narrow mass range of Axion, so it will not likely pose serious challenge for the Sgr A* spin measurement. Nevertheless, we'd like to remove this ``contamination'' for the spin; on the other hand, if the inferred dimensionless spin based on precession rate is greater than one, it is a sign of the existence of a dark matter cloud.

One may also imagine possible ways to break the degeneracy. First, if there are multiple Sgr A* EMRIs being detected at the same time, it is likely that they have different orbital radii. Therefore the relative weight expected from Lense-Thirring precession and cloud-induced precession should be different, which may be used to separately determine different components of contributions. Second, if the relevant axion mass range is ruled out by other measurements, e.g., using the spins of other massive black holes inferred from X-ray reverberation or iron line measurements~\cite{QUAX:2020adt,Gramolin:2020ict}, the possible degeneration can be obviously neglected. Third, as axion cloud may have a mass of fraction of $\alpha$ of the Sgr A* mass, it may influence the spacetime to the extent that EHT can probe the difference. We don't have a quantitative estimation for this last point, but it may worth further detailed study in the future, along with the science goals of EHT in probing dark matter.

There is also the possibility that we don't find the sign of Sgr A* EMRI using LISA, which may be explained by various reasons. In the disk migration channel, the accumulation of stellar-mass black hole in the disk close to the MBH
is a result of that the migration timescale peaks where the gravitational wave emission becomes dominant over the effect of density waves, and the number of stellar-mass BH accumulation mainly depends on the disk model and the disk lifetime. As shown in \cite{Pan2021d}, the peak of the migration timescale 
in a $\beta$ disk is rather mild and no more than two  black holes of  $10 M_\odot$ assemble in the range of $r\leq 200 M$. 
We therefore expect negligible amount of heavier BHs with mass $\geq 20 M_\odot$ considering their even lower abundance.
In a $\alpha$ disk, a larger number $(2-20)$ of $10 M_\odot$ BHs are expected to accumulate in  the range of $r\leq 200 M$
depending on the disk lifetime $(10^8-10^6)$ yr \cite{Pan2021d}. For heavier $\geq 20 M_\odot$ black holes, the number should be lower by a factor of a few again due to the lower abundance. In this case, the non-detection of Sgr A* EMRIs may indicate that the latest accretion episode is more than 1 Myrs ago such that all nearby stellar-mass black holes detectable by LISA have merged with Sgr A* due to GW emission. It is also possibly due to the fact that the accretion disk direction is random in each episode, and the lifetime of each episode is less than $10^5$ yr \cite{Pan2021d}, such that the stellar-mass black holes captured by the disk had not had enough time to migrate to the vicinity of the MBH, so the assumption of ``in-disk migration'' made in Fig.~7 of \cite{Pan2021d} breaks down. For the mass segregation scenario, it could come from the paucity of massive stellar-mass black holes within the nuclear star cluster, or other relevant assumption of the calculation that affects the final distribution in \cite{Emami2020,Emami2021}. Lastly, if there is a set of stellar-mass black holes at $\sim 10^2 M_1$ but with mass several times smaller than $20$ $M_\odot$, we may still miss the detection because of insufficient signal-to-noise ratio. 

We have assumed circular obits in the analysis as disk-driven migration efficiently damps out eccentricity, but eccentric orbits are allowed in the mass segregation scenario, or during the time that the disk has disappeared. For eccentric orbits, higher harmonics of the orbital frequency generically contribute to the waveform, which may lead to high event SNR as the frequencies of higher harmonics are closer to the sensitive band of LISA. The waveform discussed in Sec.~\ref{subsec:waveform} should include eccentricity as another parameter, and Post-Newtonian effects such as the periapsis shift should be consistently included. The accuracy in constraining various system parameters may also be improved thanks to the larger event SNR.

\acknowledgements
We thank Jun Zhang for the fruitful discussion on probing axion dark matter with Sgr A* EMRIs. We also thank Hang Yu and Zhenwei Lyu for their assistance on computing the waveform modulation due to spin induced orbital precession. S.T. thanks Zack Carson for his help in performing Fisher analyses on Mathematica. S. T., Z. P. and H. Y. are supported by the Natural Sciences and
Engineering Research Council of Canada and in part by
Perimeter Institute for Theoretical Physics. Research at
Perimeter Institute is supported in part by the Government of
Canada through the Department of Innovation, Science and
Economic Development Canada and by the Province of Ontario through the Ministry of Colleges and Universities.

\appendix

\bibliography{emri}

\begin{thebibliography}{72}%
\makeatletter
\providecommand \@ifxundefined [1]{%
 \@ifx{#1\undefined}
}%
\providecommand \@ifnum [1]{%
 \ifnum #1\expandafter \@firstoftwo
 \else \expandafter \@secondoftwo
 \fi
}%
\providecommand \@ifx [1]{%
 \ifx #1\expandafter \@firstoftwo
 \else \expandafter \@secondoftwo
 \fi
}%
\providecommand \natexlab [1]{#1}%
\providecommand \enquote  [1]{``#1''}%
\providecommand \bibnamefont  [1]{#1}%
\providecommand \bibfnamefont [1]{#1}%
\providecommand \citenamefont [1]{#1}%
\providecommand \href@noop [0]{\@secondoftwo}%
\providecommand \href [0]{\begingroup \@sanitize@url \@href}%
\providecommand \@href[1]{\@@startlink{#1}\@@href}%
\providecommand \@@href[1]{\endgroup#1\@@endlink}%
\providecommand \@sanitize@url [0]{\catcode `\\12\catcode `\$12\catcode
  `\&12\catcode `\#12\catcode `\^12\catcode `\_12\catcode `\%12\relax}%
\providecommand \@@startlink[1]{}%
\providecommand \@@endlink[0]{}%
\providecommand \url  [0]{\begingroup\@sanitize@url \@url }%
\providecommand \@url [1]{\endgroup\@href {#1}{\urlprefix }}%
\providecommand \urlprefix  [0]{URL }%
\providecommand \Eprint [0]{\href }%
\providecommand \doibase [0]{http://dx.doi.org/}%
\providecommand \selectlanguage [0]{\@gobble}%
\providecommand \bibinfo  [0]{\@secondoftwo}%
\providecommand \bibfield  [0]{\@secondoftwo}%
\providecommand \translation [1]{[#1]}%
\providecommand \BibitemOpen [0]{}%
\providecommand \bibitemStop [0]{}%
\providecommand \bibitemNoStop [0]{.\EOS\space}%
\providecommand \EOS [0]{\spacefactor3000\relax}%
\providecommand \BibitemShut  [1]{\csname bibitem#1\endcsname}%
\let\auto@bib@innerbib\@empty
\bibitem [{\citenamefont {Abbott}\ \emph
  {et~al.}(2021{\natexlab{a}})\citenamefont {Abbott} \emph
  {et~al.}}]{LIGOScientific:2021psn}%
  \BibitemOpen
  \bibfield  {author} {\bibinfo {author} {\bibfnamefont {R.}~\bibnamefont
  {Abbott}} \emph {et~al.} (\bibinfo {collaboration} {LIGO Scientific, VIRGO,
  KAGRA}),\ }\href@noop {} {\  (\bibinfo {year} {2021}{\natexlab{a}})},\
  \Eprint {http://arxiv.org/abs/2111.03634} {arXiv:2111.03634 [astro-ph.HE]}
  \BibitemShut {NoStop}%
\bibitem [{\citenamefont {Abbott}\ \emph
  {et~al.}(2021{\natexlab{b}})\citenamefont {Abbott} \emph
  {et~al.}}]{LIGOScientific:2020ibl}%
  \BibitemOpen
  \bibfield  {author} {\bibinfo {author} {\bibfnamefont {R.}~\bibnamefont
  {Abbott}} \emph {et~al.} (\bibinfo {collaboration} {LIGO Scientific,
  Virgo}),\ }\href {\doibase 10.1103/PhysRevX.11.021053} {\bibfield  {journal}
  {\bibinfo  {journal} {Phys. Rev. X}\ }\textbf {\bibinfo {volume} {11}},\
  \bibinfo {pages} {021053} (\bibinfo {year} {2021}{\natexlab{b}})},\ \Eprint
  {http://arxiv.org/abs/2010.14527} {arXiv:2010.14527 [gr-qc]} \BibitemShut
  {NoStop}%
\bibitem [{\citenamefont {Abbott}\ \emph
  {et~al.}(2021{\natexlab{c}})\citenamefont {Abbott} \emph
  {et~al.}}]{LIGOScientific:2021usb}%
  \BibitemOpen
  \bibfield  {author} {\bibinfo {author} {\bibfnamefont {R.}~\bibnamefont
  {Abbott}} \emph {et~al.} (\bibinfo {collaboration} {LIGO Scientific,
  VIRGO}),\ }\href@noop {} {\  (\bibinfo {year} {2021}{\natexlab{c}})},\
  \Eprint {http://arxiv.org/abs/2108.01045} {arXiv:2108.01045 [gr-qc]}
  \BibitemShut {NoStop}%
\bibitem [{\citenamefont {Abbott}\ \emph {et~al.}(2019)\citenamefont {Abbott}
  \emph {et~al.}}]{LIGOScientific:2018mvr}%
  \BibitemOpen
  \bibfield  {author} {\bibinfo {author} {\bibfnamefont {B.}~\bibnamefont
  {Abbott}} \emph {et~al.} (\bibinfo {collaboration} {LIGO Scientific,
  Virgo}),\ }\href {\doibase 10.1103/PhysRevX.9.031040} {\bibfield  {journal}
  {\bibinfo  {journal} {Phys. Rev. X}\ }\textbf {\bibinfo {volume} {9}},\
  \bibinfo {pages} {031040} (\bibinfo {year} {2019})},\ \Eprint
  {http://arxiv.org/abs/1811.12907} {arXiv:1811.12907 [astro-ph.HE]}
  \BibitemShut {NoStop}%
\bibitem [{\citenamefont {Akiyama}\ \emph
  {et~al.}(2019{\natexlab{a}})\citenamefont {Akiyama} \emph
  {et~al.}}]{EventHorizonTelescope:2019dse}%
  \BibitemOpen
  \bibfield  {author} {\bibinfo {author} {\bibfnamefont {K.}~\bibnamefont
  {Akiyama}} \emph {et~al.} (\bibinfo {collaboration} {Event Horizon
  Telescope}),\ }\href {\doibase 10.3847/2041-8213/ab0ec7} {\bibfield
  {journal} {\bibinfo  {journal} {Astrophys. J. Lett.}\ }\textbf {\bibinfo
  {volume} {875}},\ \bibinfo {pages} {L1} (\bibinfo {year}
  {2019}{\natexlab{a}})},\ \Eprint {http://arxiv.org/abs/1906.11238}
  {arXiv:1906.11238 [astro-ph.GA]} \BibitemShut {NoStop}%
\bibitem [{\citenamefont {Akiyama}\ \emph
  {et~al.}(2019{\natexlab{b}})\citenamefont {Akiyama} \emph
  {et~al.}}]{EventHorizonTelescope:2019ggy}%
  \BibitemOpen
  \bibfield  {author} {\bibinfo {author} {\bibfnamefont {K.}~\bibnamefont
  {Akiyama}} \emph {et~al.} (\bibinfo {collaboration} {Event Horizon
  Telescope}),\ }\href {\doibase 10.3847/2041-8213/ab1141} {\bibfield
  {journal} {\bibinfo  {journal} {Astrophys. J. Lett.}\ }\textbf {\bibinfo
  {volume} {875}},\ \bibinfo {pages} {L6} (\bibinfo {year}
  {2019}{\natexlab{b}})},\ \Eprint {http://arxiv.org/abs/1906.11243}
  {arXiv:1906.11243 [astro-ph.GA]} \BibitemShut {NoStop}%
\bibitem [{\citenamefont {Amaro-Seoane}\ \emph {et~al.}(2017)\citenamefont
  {Amaro-Seoane} \emph {et~al.}}]{Audley:2017drz}%
  \BibitemOpen
  \bibfield  {author} {\bibinfo {author} {\bibfnamefont {P.}~\bibnamefont
  {Amaro-Seoane}} \emph {et~al.} (\bibinfo {collaboration} {LISA}),\
  }\href@noop {} {\  (\bibinfo {year} {2017})},\ \Eprint
  {http://arxiv.org/abs/1702.00786} {arXiv:1702.00786 [astro-ph.IM]}
  \BibitemShut {NoStop}%
\bibitem [{\citenamefont {Luo}\ \emph {et~al.}(2016)\citenamefont {Luo} \emph
  {et~al.}}]{TianQin:2015yph}%
  \BibitemOpen
  \bibfield  {author} {\bibinfo {author} {\bibfnamefont {J.}~\bibnamefont
  {Luo}} \emph {et~al.} (\bibinfo {collaboration} {TianQin}),\ }\href {\doibase
  10.1088/0264-9381/33/3/035010} {\bibfield  {journal} {\bibinfo  {journal}
  {Class. Quant. Grav.}\ }\textbf {\bibinfo {volume} {33}},\ \bibinfo {pages}
  {035010} (\bibinfo {year} {2016})},\ \Eprint
  {http://arxiv.org/abs/1512.02076} {arXiv:1512.02076 [astro-ph.IM]}
  \BibitemShut {NoStop}%
\bibitem [{\citenamefont {Mei}\ \emph {et~al.}(2021)\citenamefont {Mei} \emph
  {et~al.}}]{TianQin:2020hid}%
  \BibitemOpen
  \bibfield  {author} {\bibinfo {author} {\bibfnamefont {J.}~\bibnamefont
  {Mei}} \emph {et~al.} (\bibinfo {collaboration} {TianQin}),\ }\href {\doibase
  10.1093/ptep/ptaa114} {\bibfield  {journal} {\bibinfo  {journal} {PTEP}\
  }\textbf {\bibinfo {volume} {2021}},\ \bibinfo {pages} {05A107} (\bibinfo
  {year} {2021})},\ \Eprint {http://arxiv.org/abs/2008.10332} {arXiv:2008.10332
  [gr-qc]} \BibitemShut {NoStop}%
\bibitem [{\citenamefont {Glampedakis}\ and\ \citenamefont
  {Babak}(2006)}]{Glampedakis:2005cf}%
  \BibitemOpen
  \bibfield  {author} {\bibinfo {author} {\bibfnamefont {K.}~\bibnamefont
  {Glampedakis}}\ and\ \bibinfo {author} {\bibfnamefont {S.}~\bibnamefont
  {Babak}},\ }\href {\doibase 10.1088/0264-9381/23/12/013} {\bibfield
  {journal} {\bibinfo  {journal} {Class. Quant. Grav.}\ }\textbf {\bibinfo
  {volume} {23}},\ \bibinfo {pages} {4167} (\bibinfo {year} {2006})},\ \Eprint
  {http://arxiv.org/abs/gr-qc/0510057} {arXiv:gr-qc/0510057} \BibitemShut
  {NoStop}%
\bibitem [{\citenamefont {Barack}\ and\ \citenamefont
  {Cutler}(2007)}]{Barack:2006pq}%
  \BibitemOpen
  \bibfield  {author} {\bibinfo {author} {\bibfnamefont {L.}~\bibnamefont
  {Barack}}\ and\ \bibinfo {author} {\bibfnamefont {C.}~\bibnamefont
  {Cutler}},\ }\href {\doibase 10.1103/PhysRevD.75.042003} {\bibfield
  {journal} {\bibinfo  {journal} {Phys. Rev. D}\ }\textbf {\bibinfo {volume}
  {75}},\ \bibinfo {pages} {042003} (\bibinfo {year} {2007})},\ \Eprint
  {http://arxiv.org/abs/gr-qc/0612029} {arXiv:gr-qc/0612029} \BibitemShut
  {NoStop}%
\bibitem [{\citenamefont {Hannuksela}\ \emph {et~al.}(2019)\citenamefont
  {Hannuksela}, \citenamefont {Wong}, \citenamefont {Brito}, \citenamefont
  {Berti},\ and\ \citenamefont {Li}}]{Hannuksela:2018izj}%
  \BibitemOpen
  \bibfield  {author} {\bibinfo {author} {\bibfnamefont {O.~A.}\ \bibnamefont
  {Hannuksela}}, \bibinfo {author} {\bibfnamefont {K.~W.~K.}\ \bibnamefont
  {Wong}}, \bibinfo {author} {\bibfnamefont {R.}~\bibnamefont {Brito}},
  \bibinfo {author} {\bibfnamefont {E.}~\bibnamefont {Berti}}, \ and\ \bibinfo
  {author} {\bibfnamefont {T.~G.~F.}\ \bibnamefont {Li}},\ }\href {\doibase
  10.1038/s41550-019-0712-4} {\bibfield  {journal} {\bibinfo  {journal} {Nature
  Astron.}\ }\textbf {\bibinfo {volume} {3}},\ \bibinfo {pages} {447} (\bibinfo
  {year} {2019})},\ \Eprint {http://arxiv.org/abs/1804.09659} {arXiv:1804.09659
  [astro-ph.HE]} \BibitemShut {NoStop}%
\bibitem [{\citenamefont {Zhang}\ and\ \citenamefont
  {Yang}(2020)}]{Zhang:2019eid}%
  \BibitemOpen
  \bibfield  {author} {\bibinfo {author} {\bibfnamefont {J.}~\bibnamefont
  {Zhang}}\ and\ \bibinfo {author} {\bibfnamefont {H.}~\bibnamefont {Yang}},\
  }\href {\doibase 10.1103/PhysRevD.101.043020} {\bibfield  {journal} {\bibinfo
   {journal} {Phys. Rev. D}\ }\textbf {\bibinfo {volume} {101}},\ \bibinfo
  {pages} {043020} (\bibinfo {year} {2020})},\ \Eprint
  {http://arxiv.org/abs/1907.13582} {arXiv:1907.13582 [gr-qc]} \BibitemShut
  {NoStop}%
\bibitem [{\citenamefont {Zhang}\ and\ \citenamefont
  {Yang}(2019)}]{Zhang:2018kib}%
  \BibitemOpen
  \bibfield  {author} {\bibinfo {author} {\bibfnamefont {J.}~\bibnamefont
  {Zhang}}\ and\ \bibinfo {author} {\bibfnamefont {H.}~\bibnamefont {Yang}},\
  }\href {\doibase 10.1103/PhysRevD.99.064018} {\bibfield  {journal} {\bibinfo
  {journal} {Phys. Rev. D}\ }\textbf {\bibinfo {volume} {99}},\ \bibinfo
  {pages} {064018} (\bibinfo {year} {2019})},\ \Eprint
  {http://arxiv.org/abs/1808.02905} {arXiv:1808.02905 [gr-qc]} \BibitemShut
  {NoStop}%
\bibitem [{\citenamefont {Bonga}\ \emph {et~al.}(2019)\citenamefont {Bonga},
  \citenamefont {Yang},\ and\ \citenamefont {Hughes}}]{Bonga:2019ycj}%
  \BibitemOpen
  \bibfield  {author} {\bibinfo {author} {\bibfnamefont {B.}~\bibnamefont
  {Bonga}}, \bibinfo {author} {\bibfnamefont {H.}~\bibnamefont {Yang}}, \ and\
  \bibinfo {author} {\bibfnamefont {S.~A.}\ \bibnamefont {Hughes}},\ }\href
  {\doibase 10.1103/PhysRevLett.123.101103} {\bibfield  {journal} {\bibinfo
  {journal} {Phys. Rev. Lett.}\ }\textbf {\bibinfo {volume} {123}},\ \bibinfo
  {pages} {101103} (\bibinfo {year} {2019})},\ \Eprint
  {http://arxiv.org/abs/1905.00030} {arXiv:1905.00030 [gr-qc]} \BibitemShut
  {NoStop}%
\bibitem [{\citenamefont {Yang}\ \emph {et~al.}(2019)\citenamefont {Yang},
  \citenamefont {Bonga}, \citenamefont {Peng},\ and\ \citenamefont
  {Li}}]{Yang:2019iqa}%
  \BibitemOpen
  \bibfield  {author} {\bibinfo {author} {\bibfnamefont {H.}~\bibnamefont
  {Yang}}, \bibinfo {author} {\bibfnamefont {B.}~\bibnamefont {Bonga}},
  \bibinfo {author} {\bibfnamefont {Z.}~\bibnamefont {Peng}}, \ and\ \bibinfo
  {author} {\bibfnamefont {G.}~\bibnamefont {Li}},\ }\href {\doibase
  10.1103/PhysRevD.100.124056} {\bibfield  {journal} {\bibinfo  {journal}
  {Phys. Rev. D}\ }\textbf {\bibinfo {volume} {100}},\ \bibinfo {pages}
  {124056} (\bibinfo {year} {2019})},\ \Eprint
  {http://arxiv.org/abs/1910.07337} {arXiv:1910.07337 [gr-qc]} \BibitemShut
  {NoStop}%
\bibitem [{\citenamefont {Yang}\ and\ \citenamefont
  {Casals}(2017)}]{Yang:2017aht}%
  \BibitemOpen
  \bibfield  {author} {\bibinfo {author} {\bibfnamefont {H.}~\bibnamefont
  {Yang}}\ and\ \bibinfo {author} {\bibfnamefont {M.}~\bibnamefont {Casals}},\
  }\href {\doibase 10.1103/PhysRevD.96.083015} {\bibfield  {journal} {\bibinfo
  {journal} {Phys. Rev. D}\ }\textbf {\bibinfo {volume} {96}},\ \bibinfo
  {pages} {083015} (\bibinfo {year} {2017})},\ \Eprint
  {http://arxiv.org/abs/1704.02022} {arXiv:1704.02022 [gr-qc]} \BibitemShut
  {NoStop}%
\bibitem [{\citenamefont {Barausse}\ \emph {et~al.}(2014)\citenamefont
  {Barausse}, \citenamefont {Cardoso},\ and\ \citenamefont
  {Pani}}]{Barausse:2014tra}%
  \BibitemOpen
  \bibfield  {author} {\bibinfo {author} {\bibfnamefont {E.}~\bibnamefont
  {Barausse}}, \bibinfo {author} {\bibfnamefont {V.}~\bibnamefont {Cardoso}}, \
  and\ \bibinfo {author} {\bibfnamefont {P.}~\bibnamefont {Pani}},\ }\href
  {\doibase 10.1103/PhysRevD.89.104059} {\bibfield  {journal} {\bibinfo
  {journal} {Phys. Rev. D}\ }\textbf {\bibinfo {volume} {89}},\ \bibinfo
  {pages} {104059} (\bibinfo {year} {2014})},\ \Eprint
  {http://arxiv.org/abs/1404.7149} {arXiv:1404.7149 [gr-qc]} \BibitemShut
  {NoStop}%
\bibitem [{\citenamefont {{Pan}}\ and\ \citenamefont
  {{Yang}}(2021)}]{Pan2021a}%
  \BibitemOpen
  \bibfield  {author} {\bibinfo {author} {\bibfnamefont {Z.}~\bibnamefont
  {{Pan}}}\ and\ \bibinfo {author} {\bibfnamefont {H.}~\bibnamefont {{Yang}}},\
  }\href {\doibase 10.1103/PhysRevD.103.103018} {\bibfield  {journal} {\bibinfo
   {journal} {\prd}\ }\textbf {\bibinfo {volume} {103}},\ \bibinfo {eid}
  {103018} (\bibinfo {year} {2021})},\ \Eprint
  {http://arxiv.org/abs/2101.09146} {arXiv:2101.09146 [astro-ph.HE]}
  \BibitemShut {NoStop}%
\bibitem [{\citenamefont {{Pan}}\ \emph
  {et~al.}(2021{\natexlab{a}})\citenamefont {{Pan}}, \citenamefont {{Lyu}},\
  and\ \citenamefont {{Yang}}}]{Pan2021b}%
  \BibitemOpen
  \bibfield  {author} {\bibinfo {author} {\bibfnamefont {Z.}~\bibnamefont
  {{Pan}}}, \bibinfo {author} {\bibfnamefont {Z.}~\bibnamefont {{Lyu}}}, \ and\
  \bibinfo {author} {\bibfnamefont {H.}~\bibnamefont {{Yang}}},\ }\href
  {\doibase 10.1103/PhysRevD.104.063007} {\bibfield  {journal} {\bibinfo
  {journal} {\prd}\ }\textbf {\bibinfo {volume} {104}},\ \bibinfo {eid}
  {063007} (\bibinfo {year} {2021}{\natexlab{a}})},\ \Eprint
  {http://arxiv.org/abs/2104.01208} {arXiv:2104.01208 [astro-ph.HE]}
  \BibitemShut {NoStop}%
\bibitem [{\citenamefont {{Amaro-Seoane}}\ and\ \citenamefont
  {{Preto}}(2011)}]{Amaro2011}%
  \BibitemOpen
  \bibfield  {author} {\bibinfo {author} {\bibfnamefont {P.}~\bibnamefont
  {{Amaro-Seoane}}}\ and\ \bibinfo {author} {\bibfnamefont {M.}~\bibnamefont
  {{Preto}}},\ }\href {\doibase 10.1088/0264-9381/28/9/094017} {\bibfield
  {journal} {\bibinfo  {journal} {Classical and Quantum Gravity}\ }\textbf
  {\bibinfo {volume} {28}},\ \bibinfo {eid} {094017} (\bibinfo {year}
  {2011})},\ \Eprint {http://arxiv.org/abs/1010.5781} {arXiv:1010.5781
  [astro-ph.CO]} \BibitemShut {NoStop}%
\bibitem [{\citenamefont {{Babak}}\ \emph {et~al.}(2017)\citenamefont
  {{Babak}}, \citenamefont {{Gair}}, \citenamefont {{Sesana}}, \citenamefont
  {{Barausse}}, \citenamefont {{Sopuerta}}, \citenamefont {{Berry}},
  \citenamefont {{Berti}}, \citenamefont {{Amaro-Seoane}}, \citenamefont
  {{Petiteau}},\ and\ \citenamefont {{Klein}}}]{Babak2017}%
  \BibitemOpen
  \bibfield  {author} {\bibinfo {author} {\bibfnamefont {S.}~\bibnamefont
  {{Babak}}}, \bibinfo {author} {\bibfnamefont {J.}~\bibnamefont {{Gair}}},
  \bibinfo {author} {\bibfnamefont {A.}~\bibnamefont {{Sesana}}}, \bibinfo
  {author} {\bibfnamefont {E.}~\bibnamefont {{Barausse}}}, \bibinfo {author}
  {\bibfnamefont {C.~F.}\ \bibnamefont {{Sopuerta}}}, \bibinfo {author}
  {\bibfnamefont {C.~P.~L.}\ \bibnamefont {{Berry}}}, \bibinfo {author}
  {\bibfnamefont {E.}~\bibnamefont {{Berti}}}, \bibinfo {author} {\bibfnamefont
  {P.}~\bibnamefont {{Amaro-Seoane}}}, \bibinfo {author} {\bibfnamefont
  {A.}~\bibnamefont {{Petiteau}}}, \ and\ \bibinfo {author} {\bibfnamefont
  {A.}~\bibnamefont {{Klein}}},\ }\href {\doibase 10.1103/PhysRevD.95.103012}
  {\bibfield  {journal} {\bibinfo  {journal} {\prd}\ }\textbf {\bibinfo
  {volume} {95}},\ \bibinfo {eid} {103012} (\bibinfo {year} {2017})},\ \Eprint
  {http://arxiv.org/abs/1703.09722} {arXiv:1703.09722 [gr-qc]} \BibitemShut
  {NoStop}%
\bibitem [{\citenamefont {{Amaro-Seoane}}(2018)}]{Amaro2018}%
  \BibitemOpen
  \bibfield  {author} {\bibinfo {author} {\bibfnamefont {P.}~\bibnamefont
  {{Amaro-Seoane}}},\ }\href {\doibase 10.1007/s41114-018-0013-8} {\bibfield
  {journal} {\bibinfo  {journal} {Living Reviews in Relativity}\ }\textbf
  {\bibinfo {volume} {21}},\ \bibinfo {eid} {4} (\bibinfo {year} {2018})},\
  \Eprint {http://arxiv.org/abs/1205.5240} {arXiv:1205.5240 [astro-ph.CO]}
  \BibitemShut {NoStop}%
\bibitem [{\citenamefont {{Amaro-Seoane}}(2020)}]{Amaro2020}%
  \BibitemOpen
  \bibfield  {author} {\bibinfo {author} {\bibfnamefont {P.}~\bibnamefont
  {{Amaro-Seoane}}},\ }\href@noop {} {\bibfield  {journal} {\bibinfo  {journal}
  {arXiv e-prints}\ ,\ \bibinfo {eid} {arXiv:2011.03059}} (\bibinfo {year}
  {2020})},\ \Eprint {http://arxiv.org/abs/2011.03059} {arXiv:2011.03059
  [gr-qc]} \BibitemShut {NoStop}%
\bibitem [{\citenamefont {Naoz}\ \emph {et~al.}(2020)\citenamefont {Naoz},
  \citenamefont {Will}, \citenamefont {Ramirez-Ruiz}, \citenamefont {Hees},
  \citenamefont {Ghez},\ and\ \citenamefont {Do}}]{Naoz:2019sjx}%
  \BibitemOpen
  \bibfield  {author} {\bibinfo {author} {\bibfnamefont {S.}~\bibnamefont
  {Naoz}}, \bibinfo {author} {\bibfnamefont {C.~M.}\ \bibnamefont {Will}},
  \bibinfo {author} {\bibfnamefont {E.}~\bibnamefont {Ramirez-Ruiz}}, \bibinfo
  {author} {\bibfnamefont {A.}~\bibnamefont {Hees}}, \bibinfo {author}
  {\bibfnamefont {A.~M.}\ \bibnamefont {Ghez}}, \ and\ \bibinfo {author}
  {\bibfnamefont {T.}~\bibnamefont {Do}},\ }\href {\doibase
  10.3847/2041-8213/ab5e3b} {\bibfield  {journal} {\bibinfo  {journal}
  {Astrophys. J. Lett.}\ }\textbf {\bibinfo {volume} {888}},\ \bibinfo {pages}
  {L8} (\bibinfo {year} {2020})},\ \Eprint {http://arxiv.org/abs/1912.04910}
  {arXiv:1912.04910 [astro-ph.GA]} \BibitemShut {NoStop}%
\bibitem [{\citenamefont {{Emami}}\ and\ \citenamefont
  {{Loeb}}(2020)}]{Emami2020}%
  \BibitemOpen
  \bibfield  {author} {\bibinfo {author} {\bibfnamefont {R.}~\bibnamefont
  {{Emami}}}\ and\ \bibinfo {author} {\bibfnamefont {A.}~\bibnamefont
  {{Loeb}}},\ }\href {\doibase 10.1088/1475-7516/2020/02/021} {\bibfield
  {journal} {\bibinfo  {journal} {jcap}\ }\textbf {\bibinfo {volume} {2020}},\
  \bibinfo {eid} {021} (\bibinfo {year} {2020})},\ \Eprint
  {http://arxiv.org/abs/1903.02578} {arXiv:1903.02578 [astro-ph.HE]}
  \BibitemShut {NoStop}%
\bibitem [{\citenamefont {{Emami}}\ and\ \citenamefont
  {{Loeb}}(2021)}]{Emami2021}%
  \BibitemOpen
  \bibfield  {author} {\bibinfo {author} {\bibfnamefont {R.}~\bibnamefont
  {{Emami}}}\ and\ \bibinfo {author} {\bibfnamefont {A.}~\bibnamefont
  {{Loeb}}},\ }\href {\doibase 10.1093/mnras/stab290} {\bibfield  {journal}
  {\bibinfo  {journal} {mnras}\ }\textbf {\bibinfo {volume} {502}},\ \bibinfo
  {pages} {3932} (\bibinfo {year} {2021})},\ \Eprint
  {http://arxiv.org/abs/1903.02579} {arXiv:1903.02579 [astro-ph.HE]}
  \BibitemShut {NoStop}%
\bibitem [{\citenamefont {{Levin}}\ and\ \citenamefont
  {{Beloborodov}}(2003)}]{Levin2003}%
  \BibitemOpen
  \bibfield  {author} {\bibinfo {author} {\bibfnamefont {Y.}~\bibnamefont
  {{Levin}}}\ and\ \bibinfo {author} {\bibfnamefont {A.~M.}\ \bibnamefont
  {{Beloborodov}}},\ }\href {\doibase 10.1086/376675} {\bibfield  {journal}
  {\bibinfo  {journal} {apjl}\ }\textbf {\bibinfo {volume} {590}},\ \bibinfo
  {pages} {L33} (\bibinfo {year} {2003})},\ \Eprint
  {http://arxiv.org/abs/astro-ph/0303436} {arXiv:astro-ph/0303436 [astro-ph]}
  \BibitemShut {NoStop}%
\bibitem [{\citenamefont {{Gilbaum}}\ and\ \citenamefont
  {{Stone}}(2021)}]{Gilbaum2021}%
  \BibitemOpen
  \bibfield  {author} {\bibinfo {author} {\bibfnamefont {S.}~\bibnamefont
  {{Gilbaum}}}\ and\ \bibinfo {author} {\bibfnamefont {N.~C.}\ \bibnamefont
  {{Stone}}},\ }\href@noop {} {\bibfield  {journal} {\bibinfo  {journal} {arXiv
  e-prints}\ ,\ \bibinfo {eid} {arXiv:2107.07519}} (\bibinfo {year} {2021})},\
  \Eprint {http://arxiv.org/abs/2107.07519} {arXiv:2107.07519 [astro-ph.HE]}
  \BibitemShut {NoStop}%
\bibitem [{\citenamefont {Pan}\ and\ \citenamefont {Yang}(2021)}]{Pan2021c}%
  \BibitemOpen
  \bibfield  {author} {\bibinfo {author} {\bibfnamefont {Z.}~\bibnamefont
  {Pan}}\ and\ \bibinfo {author} {\bibfnamefont {H.}~\bibnamefont {Yang}},\
  }\href {\doibase 10.3847/1538-4357/ac249c} {\bibfield  {journal} {\bibinfo
  {journal} {The Astrophysical Journal}\ }\textbf {\bibinfo {volume} {923}},\
  \bibinfo {pages} {173} (\bibinfo {year} {2021})}\BibitemShut {NoStop}%
\bibitem [{\citenamefont {{Pan}}\ \emph
  {et~al.}(2021{\natexlab{b}})\citenamefont {{Pan}}, \citenamefont {{Lyu}},\
  and\ \citenamefont {{Yang}}}]{Pan2021d}%
  \BibitemOpen
  \bibfield  {author} {\bibinfo {author} {\bibfnamefont {Z.}~\bibnamefont
  {{Pan}}}, \bibinfo {author} {\bibfnamefont {Z.}~\bibnamefont {{Lyu}}}, \ and\
  \bibinfo {author} {\bibfnamefont {H.}~\bibnamefont {{Yang}}},\ }\href@noop {}
  {\bibfield  {journal} {\bibinfo  {journal} {arXiv e-prints}\ ,\ \bibinfo
  {eid} {arXiv:2112.10237}} (\bibinfo {year} {2021}{\natexlab{b}})},\ \Eprint
  {http://arxiv.org/abs/2112.10237} {arXiv:2112.10237 [astro-ph.HE]}
  \BibitemShut {NoStop}%
\bibitem [{\citenamefont {Gourgoulhon}\ \emph {et~al.}(2019)\citenamefont
  {Gourgoulhon}, \citenamefont {Le~Tiec}, \citenamefont {Vincent},\ and\
  \citenamefont {Warburton}}]{Gourgoulhon:2019iyu}%
  \BibitemOpen
  \bibfield  {author} {\bibinfo {author} {\bibfnamefont {E.}~\bibnamefont
  {Gourgoulhon}}, \bibinfo {author} {\bibfnamefont {A.}~\bibnamefont
  {Le~Tiec}}, \bibinfo {author} {\bibfnamefont {F.~H.}\ \bibnamefont
  {Vincent}}, \ and\ \bibinfo {author} {\bibfnamefont {N.}~\bibnamefont
  {Warburton}},\ }\href {\doibase 10.1051/0004-6361/201935406} {\bibfield
  {journal} {\bibinfo  {journal} {Astron. Astrophys.}\ }\textbf {\bibinfo
  {volume} {627}},\ \bibinfo {pages} {A92} (\bibinfo {year} {2019})},\ \Eprint
  {http://arxiv.org/abs/1903.02049} {arXiv:1903.02049 [gr-qc]} \BibitemShut
  {NoStop}%
\bibitem [{\citenamefont {Tamburini}\ \emph {et~al.}(2020)\citenamefont
  {Tamburini}, \citenamefont {Thid\'e},\ and\ \citenamefont
  {Della~Valle}}]{Tamburini:2019vrf}%
  \BibitemOpen
  \bibfield  {author} {\bibinfo {author} {\bibfnamefont {F.}~\bibnamefont
  {Tamburini}}, \bibinfo {author} {\bibfnamefont {B.}~\bibnamefont {Thid\'e}},
  \ and\ \bibinfo {author} {\bibfnamefont {M.}~\bibnamefont {Della~Valle}},\
  }\href {\doibase 10.1093/mnrasl/slz176} {\bibfield  {journal} {\bibinfo
  {journal} {Mon. Not. Roy. Astron. Soc.}\ }\textbf {\bibinfo {volume} {492}},\
  \bibinfo {pages} {L22} (\bibinfo {year} {2020})},\ \Eprint
  {http://arxiv.org/abs/1904.07923} {arXiv:1904.07923 [astro-ph.HE]}
  \BibitemShut {NoStop}%
\bibitem [{\citenamefont {Bambi}\ \emph {et~al.}(2019)\citenamefont {Bambi},
  \citenamefont {Freese}, \citenamefont {Vagnozzi},\ and\ \citenamefont
  {Visinelli}}]{Bambi:2019tjh}%
  \BibitemOpen
  \bibfield  {author} {\bibinfo {author} {\bibfnamefont {C.}~\bibnamefont
  {Bambi}}, \bibinfo {author} {\bibfnamefont {K.}~\bibnamefont {Freese}},
  \bibinfo {author} {\bibfnamefont {S.}~\bibnamefont {Vagnozzi}}, \ and\
  \bibinfo {author} {\bibfnamefont {L.}~\bibnamefont {Visinelli}},\ }\href
  {\doibase 10.1103/PhysRevD.100.044057} {\bibfield  {journal} {\bibinfo
  {journal} {Phys. Rev. D}\ }\textbf {\bibinfo {volume} {100}},\ \bibinfo
  {pages} {044057} (\bibinfo {year} {2019})},\ \Eprint
  {http://arxiv.org/abs/1904.12983} {arXiv:1904.12983 [gr-qc]} \BibitemShut
  {NoStop}%
\bibitem [{\citenamefont {Cutler}(1998)}]{Cutler:1997ta}%
  \BibitemOpen
  \bibfield  {author} {\bibinfo {author} {\bibfnamefont {C.}~\bibnamefont
  {Cutler}},\ }\href {\doibase 10.1103/PhysRevD.57.7089} {\bibfield  {journal}
  {\bibinfo  {journal} {Phys. Rev. D}\ }\textbf {\bibinfo {volume} {57}},\
  \bibinfo {pages} {7089} (\bibinfo {year} {1998})},\ \Eprint
  {http://arxiv.org/abs/gr-qc/9703068} {arXiv:gr-qc/9703068} \BibitemShut
  {NoStop}%
\bibitem [{\citenamefont {Apostolatos}\ \emph {et~al.}(1994)\citenamefont
  {Apostolatos}, \citenamefont {Cutler}, \citenamefont {Sussman},\ and\
  \citenamefont {Thorne}}]{Apostolatos:1994mx}%
  \BibitemOpen
  \bibfield  {author} {\bibinfo {author} {\bibfnamefont {T.~A.}\ \bibnamefont
  {Apostolatos}}, \bibinfo {author} {\bibfnamefont {C.}~\bibnamefont {Cutler}},
  \bibinfo {author} {\bibfnamefont {G.~J.}\ \bibnamefont {Sussman}}, \ and\
  \bibinfo {author} {\bibfnamefont {K.~S.}\ \bibnamefont {Thorne}},\ }\href
  {\doibase 10.1103/PhysRevD.49.6274} {\bibfield  {journal} {\bibinfo
  {journal} {Phys. Rev. D}\ }\textbf {\bibinfo {volume} {49}},\ \bibinfo
  {pages} {6274} (\bibinfo {year} {1994})}\BibitemShut {NoStop}%
\bibitem [{\citenamefont {Abuter}\ \emph {et~al.}(2021)\citenamefont {Abuter}
  \emph {et~al.}}]{GRAVITY:2021xju}%
  \BibitemOpen
  \bibfield  {author} {\bibinfo {author} {\bibfnamefont {R.}~\bibnamefont
  {Abuter}} \emph {et~al.} (\bibinfo {collaboration} {GRAVITY}),\ }\href
  {\doibase 10.1051/0004-6361/202142465} {\  (\bibinfo {year} {2021}),\
  10.1051/0004-6361/202142465},\ \Eprint {http://arxiv.org/abs/2112.07478}
  {arXiv:2112.07478 [astro-ph.GA]} \BibitemShut {NoStop}%
\bibitem [{\citenamefont {Abuter}\ \emph {et~al.}(2020)\citenamefont {Abuter}
  \emph {et~al.}}]{GRAVITY:2020gka}%
  \BibitemOpen
  \bibfield  {author} {\bibinfo {author} {\bibfnamefont {R.}~\bibnamefont
  {Abuter}} \emph {et~al.} (\bibinfo {collaboration} {GRAVITY}),\ }\href
  {\doibase 10.1051/0004-6361/202037813} {\bibfield  {journal} {\bibinfo
  {journal} {Astron. Astrophys.}\ }\textbf {\bibinfo {volume} {636}},\ \bibinfo
  {pages} {L5} (\bibinfo {year} {2020})},\ \Eprint
  {http://arxiv.org/abs/2004.07187} {arXiv:2004.07187 [astro-ph.GA]}
  \BibitemShut {NoStop}%
\bibitem [{\citenamefont {Abuter}\ \emph {et~al.}(2018)\citenamefont {Abuter},
  \citenamefont {Amorim}, \citenamefont {Bauböck}, \citenamefont {Berger},
  \citenamefont {Bonnet}, \citenamefont {Brandner}, \citenamefont {Clénet},
  \citenamefont {Coudé~du Foresto}, \citenamefont {de~Zeeuw},\ and\
  \citenamefont {et~al.}}]{2018}%
  \BibitemOpen
  \bibfield  {author} {\bibinfo {author} {\bibfnamefont {R.}~\bibnamefont
  {Abuter}}, \bibinfo {author} {\bibfnamefont {A.}~\bibnamefont {Amorim}},
  \bibinfo {author} {\bibfnamefont {M.}~\bibnamefont {Bauböck}}, \bibinfo
  {author} {\bibfnamefont {J.~P.}\ \bibnamefont {Berger}}, \bibinfo {author}
  {\bibfnamefont {H.}~\bibnamefont {Bonnet}}, \bibinfo {author} {\bibfnamefont
  {W.}~\bibnamefont {Brandner}}, \bibinfo {author} {\bibfnamefont
  {Y.}~\bibnamefont {Clénet}}, \bibinfo {author} {\bibfnamefont
  {V.}~\bibnamefont {Coudé~du Foresto}}, \bibinfo {author} {\bibfnamefont
  {P.~T.}\ \bibnamefont {de~Zeeuw}}, \ and\ \bibinfo {author} {\bibnamefont
  {et~al.}},\ }\href {\doibase 10.1051/0004-6361/201834294} {\bibfield
  {journal} {\bibinfo  {journal} {Astronomy \& Astrophysics}\ }\textbf
  {\bibinfo {volume} {618}},\ \bibinfo {pages} {L10} (\bibinfo {year}
  {2018})}\BibitemShut {NoStop}%
\bibitem [{\citenamefont {Robson}\ \emph {et~al.}(2019)\citenamefont {Robson},
  \citenamefont {Cornish},\ and\ \citenamefont {Liu}}]{Robson:2018ifk}%
  \BibitemOpen
  \bibfield  {author} {\bibinfo {author} {\bibfnamefont {T.}~\bibnamefont
  {Robson}}, \bibinfo {author} {\bibfnamefont {N.~J.}\ \bibnamefont {Cornish}},
  \ and\ \bibinfo {author} {\bibfnamefont {C.}~\bibnamefont {Liu}},\ }\href
  {\doibase 10.1088/1361-6382/ab1101} {\bibfield  {journal} {\bibinfo
  {journal} {Class. Quant. Grav.}\ }\textbf {\bibinfo {volume} {36}},\ \bibinfo
  {pages} {105011} (\bibinfo {year} {2019})},\ \Eprint
  {http://arxiv.org/abs/1803.01944} {arXiv:1803.01944 [astro-ph.HE]}
  \BibitemShut {NoStop}%
\bibitem [{\citenamefont {Moore}\ \emph {et~al.}(2019)\citenamefont {Moore},
  \citenamefont {Gerosa},\ and\ \citenamefont {Klein}}]{Moore:2019pke}%
  \BibitemOpen
  \bibfield  {author} {\bibinfo {author} {\bibfnamefont {C.~J.}\ \bibnamefont
  {Moore}}, \bibinfo {author} {\bibfnamefont {D.}~\bibnamefont {Gerosa}}, \
  and\ \bibinfo {author} {\bibfnamefont {A.}~\bibnamefont {Klein}},\ }\href
  {\doibase 10.1093/mnrasl/slz104} {\bibfield  {journal} {\bibinfo  {journal}
  {Mon. Not. Roy. Astron. Soc.}\ }\textbf {\bibinfo {volume} {488}},\ \bibinfo
  {pages} {L94} (\bibinfo {year} {2019})},\ \Eprint
  {http://arxiv.org/abs/1905.11998} {arXiv:1905.11998 [astro-ph.HE]}
  \BibitemShut {NoStop}%
\bibitem [{\citenamefont {Abbott}\ \emph {et~al.}(2016)\citenamefont {Abbott}
  \emph {et~al.}}]{LIGOScientific:2016dsl}%
  \BibitemOpen
  \bibfield  {author} {\bibinfo {author} {\bibfnamefont {B.~P.}\ \bibnamefont
  {Abbott}} \emph {et~al.} (\bibinfo {collaboration} {LIGO Scientific,
  Virgo}),\ }\href {\doibase 10.1103/PhysRevX.6.041015} {\bibfield  {journal}
  {\bibinfo  {journal} {Phys. Rev. X}\ }\textbf {\bibinfo {volume} {6}},\
  \bibinfo {pages} {041015} (\bibinfo {year} {2016})},\ \bibinfo {note}
  {[Erratum: Phys.Rev.X 8, 039903 (2018)]},\ \Eprint
  {http://arxiv.org/abs/1606.04856} {arXiv:1606.04856 [gr-qc]} \BibitemShut
  {NoStop}%
\bibitem [{\citenamefont {Cornish}\ and\ \citenamefont
  {Porter}(2005)}]{Cornish:2005hd}%
  \BibitemOpen
  \bibfield  {author} {\bibinfo {author} {\bibfnamefont {N.~J.}\ \bibnamefont
  {Cornish}}\ and\ \bibinfo {author} {\bibfnamefont {E.~K.}\ \bibnamefont
  {Porter}},\ }\href {\doibase 10.1088/0264-9381/22/18/S06} {\bibfield
  {journal} {\bibinfo  {journal} {Class. Quant. Grav.}\ }\textbf {\bibinfo
  {volume} {22}},\ \bibinfo {pages} {S927} (\bibinfo {year} {2005})},\ \Eprint
  {http://arxiv.org/abs/gr-qc/0504012} {arXiv:gr-qc/0504012} \BibitemShut
  {NoStop}%
\bibitem [{\citenamefont {Gondolo}\ and\ \citenamefont
  {Silk}(1999)}]{Gondolo:1999ef}%
  \BibitemOpen
  \bibfield  {author} {\bibinfo {author} {\bibfnamefont {P.}~\bibnamefont
  {Gondolo}}\ and\ \bibinfo {author} {\bibfnamefont {J.}~\bibnamefont {Silk}},\
  }\href {\doibase 10.1103/PhysRevLett.83.1719} {\bibfield  {journal} {\bibinfo
   {journal} {Phys. Rev. Lett.}\ }\textbf {\bibinfo {volume} {83}},\ \bibinfo
  {pages} {1719} (\bibinfo {year} {1999})},\ \Eprint
  {http://arxiv.org/abs/astro-ph/9906391} {arXiv:astro-ph/9906391} \BibitemShut
  {NoStop}%
\bibitem [{\citenamefont {Ullio}\ \emph {et~al.}(2001)\citenamefont {Ullio},
  \citenamefont {Zhao},\ and\ \citenamefont {Kamionkowski}}]{Ullio:2001fb}%
  \BibitemOpen
  \bibfield  {author} {\bibinfo {author} {\bibfnamefont {P.}~\bibnamefont
  {Ullio}}, \bibinfo {author} {\bibfnamefont {H.}~\bibnamefont {Zhao}}, \ and\
  \bibinfo {author} {\bibfnamefont {M.}~\bibnamefont {Kamionkowski}},\ }\href
  {\doibase 10.1103/PhysRevD.64.043504} {\bibfield  {journal} {\bibinfo
  {journal} {Phys. Rev. D}\ }\textbf {\bibinfo {volume} {64}},\ \bibinfo
  {pages} {043504} (\bibinfo {year} {2001})},\ \Eprint
  {http://arxiv.org/abs/astro-ph/0101481} {arXiv:astro-ph/0101481} \BibitemShut
  {NoStop}%
\bibitem [{\citenamefont {Kavanagh}\ \emph {et~al.}(2020)\citenamefont
  {Kavanagh}, \citenamefont {Nichols}, \citenamefont {Bertone},\ and\
  \citenamefont {Gaggero}}]{Kavanagh:2020cfn}%
  \BibitemOpen
  \bibfield  {author} {\bibinfo {author} {\bibfnamefont {B.~J.}\ \bibnamefont
  {Kavanagh}}, \bibinfo {author} {\bibfnamefont {D.~A.}\ \bibnamefont
  {Nichols}}, \bibinfo {author} {\bibfnamefont {G.}~\bibnamefont {Bertone}}, \
  and\ \bibinfo {author} {\bibfnamefont {D.}~\bibnamefont {Gaggero}},\ }\href
  {\doibase 10.1103/PhysRevD.102.083006} {\bibfield  {journal} {\bibinfo
  {journal} {Phys. Rev. D}\ }\textbf {\bibinfo {volume} {102}},\ \bibinfo
  {pages} {083006} (\bibinfo {year} {2020})},\ \Eprint
  {http://arxiv.org/abs/2002.12811} {arXiv:2002.12811 [gr-qc]} \BibitemShut
  {NoStop}%
\bibitem [{\citenamefont {Coogan}\ \emph {et~al.}(2021)\citenamefont {Coogan},
  \citenamefont {Bertone}, \citenamefont {Gaggero}, \citenamefont {Kavanagh},\
  and\ \citenamefont {Nichols}}]{Coogan:2021uqv}%
  \BibitemOpen
  \bibfield  {author} {\bibinfo {author} {\bibfnamefont {A.}~\bibnamefont
  {Coogan}}, \bibinfo {author} {\bibfnamefont {G.}~\bibnamefont {Bertone}},
  \bibinfo {author} {\bibfnamefont {D.}~\bibnamefont {Gaggero}}, \bibinfo
  {author} {\bibfnamefont {B.~J.}\ \bibnamefont {Kavanagh}}, \ and\ \bibinfo
  {author} {\bibfnamefont {D.~A.}\ \bibnamefont {Nichols}},\ }\href@noop {} {\
  (\bibinfo {year} {2021})},\ \Eprint {http://arxiv.org/abs/2108.04154}
  {arXiv:2108.04154 [gr-qc]} \BibitemShut {NoStop}%
\bibitem [{\citenamefont {Eda}\ \emph {et~al.}(2015)\citenamefont {Eda},
  \citenamefont {Itoh}, \citenamefont {Kuroyanagi},\ and\ \citenamefont
  {Silk}}]{Eda:2014kra}%
  \BibitemOpen
  \bibfield  {author} {\bibinfo {author} {\bibfnamefont {K.}~\bibnamefont
  {Eda}}, \bibinfo {author} {\bibfnamefont {Y.}~\bibnamefont {Itoh}}, \bibinfo
  {author} {\bibfnamefont {S.}~\bibnamefont {Kuroyanagi}}, \ and\ \bibinfo
  {author} {\bibfnamefont {J.}~\bibnamefont {Silk}},\ }\href {\doibase
  10.1103/PhysRevD.91.044045} {\bibfield  {journal} {\bibinfo  {journal} {Phys.
  Rev. D}\ }\textbf {\bibinfo {volume} {91}},\ \bibinfo {pages} {044045}
  (\bibinfo {year} {2015})},\ \Eprint {http://arxiv.org/abs/1408.3534}
  {arXiv:1408.3534 [gr-qc]} \BibitemShut {NoStop}%
\bibitem [{\citenamefont {Navarro}\ \emph {et~al.}(1996)\citenamefont
  {Navarro}, \citenamefont {Frenk},\ and\ \citenamefont
  {White}}]{Navarro:1995iw}%
  \BibitemOpen
  \bibfield  {author} {\bibinfo {author} {\bibfnamefont {J.~F.}\ \bibnamefont
  {Navarro}}, \bibinfo {author} {\bibfnamefont {C.~S.}\ \bibnamefont {Frenk}},
  \ and\ \bibinfo {author} {\bibfnamefont {S.~D.~M.}\ \bibnamefont {White}},\
  }\href {\doibase 10.1086/177173} {\bibfield  {journal} {\bibinfo  {journal}
  {Astrophys. J.}\ }\textbf {\bibinfo {volume} {462}},\ \bibinfo {pages} {563}
  (\bibinfo {year} {1996})},\ \Eprint {http://arxiv.org/abs/astro-ph/9508025}
  {arXiv:astro-ph/9508025} \BibitemShut {NoStop}%
\bibitem [{\citenamefont {Dutton}\ and\ \citenamefont
  {Macci\`o}(2014)}]{Dutton:2014xda}%
  \BibitemOpen
  \bibfield  {author} {\bibinfo {author} {\bibfnamefont {A.~A.}\ \bibnamefont
  {Dutton}}\ and\ \bibinfo {author} {\bibfnamefont {A.~V.}\ \bibnamefont
  {Macci\`o}},\ }\href {\doibase 10.1093/mnras/stu742} {\bibfield  {journal}
  {\bibinfo  {journal} {Mon. Not. Roy. Astron. Soc.}\ }\textbf {\bibinfo
  {volume} {441}},\ \bibinfo {pages} {3359} (\bibinfo {year} {2014})},\ \Eprint
  {http://arxiv.org/abs/1402.7073} {arXiv:1402.7073 [astro-ph.CO]} \BibitemShut
  {NoStop}%
\bibitem [{\citenamefont {Ferrarese}(2002)}]{Ferrarese:2002ct}%
  \BibitemOpen
  \bibfield  {author} {\bibinfo {author} {\bibfnamefont {L.}~\bibnamefont
  {Ferrarese}},\ }\href {\doibase 10.1086/342308} {\bibfield  {journal}
  {\bibinfo  {journal} {Astrophys. J.}\ }\textbf {\bibinfo {volume} {578}},\
  \bibinfo {pages} {90} (\bibinfo {year} {2002})},\ \Eprint
  {http://arxiv.org/abs/astro-ph/0203469} {arXiv:astro-ph/0203469} \BibitemShut
  {NoStop}%
\bibitem [{\citenamefont {Merritt}(2004)}]{Merritt:2003qk}%
  \BibitemOpen
  \bibfield  {author} {\bibinfo {author} {\bibfnamefont {D.}~\bibnamefont
  {Merritt}},\ }\href {\doibase 10.1103/PhysRevLett.92.201304} {\bibfield
  {journal} {\bibinfo  {journal} {Phys. Rev. Lett.}\ }\textbf {\bibinfo
  {volume} {92}},\ \bibinfo {pages} {201304} (\bibinfo {year} {2004})},\
  \Eprint {http://arxiv.org/abs/astro-ph/0311594} {arXiv:astro-ph/0311594}
  \BibitemShut {NoStop}%
\bibitem [{\citenamefont {Marsh}(2016)}]{Marsh:2015xka}%
  \BibitemOpen
  \bibfield  {author} {\bibinfo {author} {\bibfnamefont {D.~J.~E.}\
  \bibnamefont {Marsh}},\ }\href {\doibase 10.1016/j.physrep.2016.06.005}
  {\bibfield  {journal} {\bibinfo  {journal} {Phys. Rept.}\ }\textbf {\bibinfo
  {volume} {643}},\ \bibinfo {pages} {1} (\bibinfo {year} {2016})},\ \Eprint
  {http://arxiv.org/abs/1510.07633} {arXiv:1510.07633 [astro-ph.CO]}
  \BibitemShut {NoStop}%
\bibitem [{\citenamefont {Arvanitaki}\ \emph {et~al.}(2010)\citenamefont
  {Arvanitaki}, \citenamefont {Dimopoulos}, \citenamefont {Dubovsky},
  \citenamefont {Kaloper},\ and\ \citenamefont
  {March-Russell}}]{Arvanitaki:2009fg}%
  \BibitemOpen
  \bibfield  {author} {\bibinfo {author} {\bibfnamefont {A.}~\bibnamefont
  {Arvanitaki}}, \bibinfo {author} {\bibfnamefont {S.}~\bibnamefont
  {Dimopoulos}}, \bibinfo {author} {\bibfnamefont {S.}~\bibnamefont
  {Dubovsky}}, \bibinfo {author} {\bibfnamefont {N.}~\bibnamefont {Kaloper}}, \
  and\ \bibinfo {author} {\bibfnamefont {J.}~\bibnamefont {March-Russell}},\
  }\href {\doibase 10.1103/PhysRevD.81.123530} {\bibfield  {journal} {\bibinfo
  {journal} {Phys. Rev. D}\ }\textbf {\bibinfo {volume} {81}},\ \bibinfo
  {pages} {123530} (\bibinfo {year} {2010})},\ \Eprint
  {http://arxiv.org/abs/0905.4720} {arXiv:0905.4720 [hep-th]} \BibitemShut
  {NoStop}%
\bibitem [{\citenamefont {Peccei}\ and\ \citenamefont
  {Quinn}(1977)}]{Peccei:1977hh}%
  \BibitemOpen
  \bibfield  {author} {\bibinfo {author} {\bibfnamefont {R.~D.}\ \bibnamefont
  {Peccei}}\ and\ \bibinfo {author} {\bibfnamefont {H.~R.}\ \bibnamefont
  {Quinn}},\ }\href {\doibase 10.1103/PhysRevLett.38.1440} {\bibfield
  {journal} {\bibinfo  {journal} {Phys. Rev. Lett.}\ }\textbf {\bibinfo
  {volume} {38}},\ \bibinfo {pages} {1440} (\bibinfo {year}
  {1977})}\BibitemShut {NoStop}%
\bibitem [{\citenamefont {Weinberg}(1978)}]{Weinberg:1977ma}%
  \BibitemOpen
  \bibfield  {author} {\bibinfo {author} {\bibfnamefont {S.}~\bibnamefont
  {Weinberg}},\ }\href {\doibase 10.1103/PhysRevLett.40.223} {\bibfield
  {journal} {\bibinfo  {journal} {Phys. Rev. Lett.}\ }\textbf {\bibinfo
  {volume} {40}},\ \bibinfo {pages} {223} (\bibinfo {year} {1978})}\BibitemShut
  {NoStop}%
\bibitem [{\citenamefont {Dine}\ and\ \citenamefont
  {Fischler}(1983)}]{Dine:1982ah}%
  \BibitemOpen
  \bibfield  {author} {\bibinfo {author} {\bibfnamefont {M.}~\bibnamefont
  {Dine}}\ and\ \bibinfo {author} {\bibfnamefont {W.}~\bibnamefont
  {Fischler}},\ }\href {\doibase 10.1016/0370-2693(83)90639-1} {\bibfield
  {journal} {\bibinfo  {journal} {Phys. Lett. B}\ }\textbf {\bibinfo {volume}
  {120}},\ \bibinfo {pages} {137} (\bibinfo {year} {1983})}\BibitemShut
  {NoStop}%
\bibitem [{\citenamefont {Abbott}\ and\ \citenamefont
  {Sikivie}(1983)}]{Abbott:1982af}%
  \BibitemOpen
  \bibfield  {author} {\bibinfo {author} {\bibfnamefont {L.~F.}\ \bibnamefont
  {Abbott}}\ and\ \bibinfo {author} {\bibfnamefont {P.}~\bibnamefont
  {Sikivie}},\ }\href {\doibase 10.1016/0370-2693(83)90638-X} {\bibfield
  {journal} {\bibinfo  {journal} {Phys. Lett. B}\ }\textbf {\bibinfo {volume}
  {120}},\ \bibinfo {pages} {133} (\bibinfo {year} {1983})}\BibitemShut
  {NoStop}%
\bibitem [{\citenamefont {Press}\ \emph {et~al.}(1990)\citenamefont {Press},
  \citenamefont {Ryden},\ and\ \citenamefont {Spergel}}]{Press:1989id}%
  \BibitemOpen
  \bibfield  {author} {\bibinfo {author} {\bibfnamefont {W.~H.}\ \bibnamefont
  {Press}}, \bibinfo {author} {\bibfnamefont {B.~S.}\ \bibnamefont {Ryden}}, \
  and\ \bibinfo {author} {\bibfnamefont {D.~N.}\ \bibnamefont {Spergel}},\
  }\href {\doibase 10.1103/PhysRevLett.64.1084} {\bibfield  {journal} {\bibinfo
   {journal} {Phys. Rev. Lett.}\ }\textbf {\bibinfo {volume} {64}},\ \bibinfo
  {pages} {1084} (\bibinfo {year} {1990})}\BibitemShut {NoStop}%
\bibitem [{\citenamefont {Peebles}(2000)}]{Peebles:2000yy}%
  \BibitemOpen
  \bibfield  {author} {\bibinfo {author} {\bibfnamefont {P.~J.~E.}\
  \bibnamefont {Peebles}},\ }\href {\doibase 10.1086/312677} {\bibfield
  {journal} {\bibinfo  {journal} {Astrophys. J. Lett.}\ }\textbf {\bibinfo
  {volume} {534}},\ \bibinfo {pages} {L127} (\bibinfo {year} {2000})},\ \Eprint
  {http://arxiv.org/abs/astro-ph/0002495} {arXiv:astro-ph/0002495} \BibitemShut
  {NoStop}%
\bibitem [{\citenamefont {Amendola}\ and\ \citenamefont
  {Barbieri}(2006)}]{Amendola:2005ad}%
  \BibitemOpen
  \bibfield  {author} {\bibinfo {author} {\bibfnamefont {L.}~\bibnamefont
  {Amendola}}\ and\ \bibinfo {author} {\bibfnamefont {R.}~\bibnamefont
  {Barbieri}},\ }\href {\doibase 10.1016/j.physletb.2006.08.069} {\bibfield
  {journal} {\bibinfo  {journal} {Phys. Lett. B}\ }\textbf {\bibinfo {volume}
  {642}},\ \bibinfo {pages} {192} (\bibinfo {year} {2006})},\ \Eprint
  {http://arxiv.org/abs/hep-ph/0509257} {arXiv:hep-ph/0509257} \BibitemShut
  {NoStop}%
\bibitem [{\citenamefont {Hui}\ \emph {et~al.}(2017)\citenamefont {Hui},
  \citenamefont {Ostriker}, \citenamefont {Tremaine},\ and\ \citenamefont
  {Witten}}]{Hui:2016ltb}%
  \BibitemOpen
  \bibfield  {author} {\bibinfo {author} {\bibfnamefont {L.}~\bibnamefont
  {Hui}}, \bibinfo {author} {\bibfnamefont {J.~P.}\ \bibnamefont {Ostriker}},
  \bibinfo {author} {\bibfnamefont {S.}~\bibnamefont {Tremaine}}, \ and\
  \bibinfo {author} {\bibfnamefont {E.}~\bibnamefont {Witten}},\ }\href
  {\doibase 10.1103/PhysRevD.95.043541} {\bibfield  {journal} {\bibinfo
  {journal} {Phys. Rev. D}\ }\textbf {\bibinfo {volume} {95}},\ \bibinfo
  {pages} {043541} (\bibinfo {year} {2017})},\ \Eprint
  {http://arxiv.org/abs/1610.08297} {arXiv:1610.08297 [astro-ph.CO]}
  \BibitemShut {NoStop}%
\bibitem [{\citenamefont {Preskill}\ \emph {et~al.}(1983)\citenamefont
  {Preskill}, \citenamefont {Wise},\ and\ \citenamefont
  {Wilczek}}]{Preskill:1982cy}%
  \BibitemOpen
  \bibfield  {author} {\bibinfo {author} {\bibfnamefont {J.}~\bibnamefont
  {Preskill}}, \bibinfo {author} {\bibfnamefont {M.~B.}\ \bibnamefont {Wise}},
  \ and\ \bibinfo {author} {\bibfnamefont {F.}~\bibnamefont {Wilczek}},\ }\href
  {\doibase 10.1016/0370-2693(83)90637-8} {\bibfield  {journal} {\bibinfo
  {journal} {Phys. Lett. B}\ }\textbf {\bibinfo {volume} {120}},\ \bibinfo
  {pages} {127} (\bibinfo {year} {1983})}\BibitemShut {NoStop}%
\bibitem [{\citenamefont {Baumann}\ \emph {et~al.}(2019)\citenamefont
  {Baumann}, \citenamefont {Chia},\ and\ \citenamefont
  {Porto}}]{Baumann:2018vus}%
  \BibitemOpen
  \bibfield  {author} {\bibinfo {author} {\bibfnamefont {D.}~\bibnamefont
  {Baumann}}, \bibinfo {author} {\bibfnamefont {H.~S.}\ \bibnamefont {Chia}}, \
  and\ \bibinfo {author} {\bibfnamefont {R.~A.}\ \bibnamefont {Porto}},\ }\href
  {\doibase 10.1103/PhysRevD.99.044001} {\bibfield  {journal} {\bibinfo
  {journal} {Phys. Rev. D}\ }\textbf {\bibinfo {volume} {99}},\ \bibinfo
  {pages} {044001} (\bibinfo {year} {2019})},\ \Eprint
  {http://arxiv.org/abs/1804.03208} {arXiv:1804.03208 [gr-qc]} \BibitemShut
  {NoStop}%
\bibitem [{\citenamefont {Poisson}\ and\ \citenamefont
  {Will}(2014)}]{2014grav.bookpoissonwill}%
  \BibitemOpen
  \bibfield  {author} {\bibinfo {author} {\bibfnamefont {E.}~\bibnamefont
  {Poisson}}\ and\ \bibinfo {author} {\bibfnamefont {C.~M.}\ \bibnamefont
  {Will}},\ }\href {\doibase 10.1017/CBO9781139507486} {\emph {\bibinfo {title}
  {Gravity: Newtonian, Post-Newtonian, Relativistic}}}\ (\bibinfo  {publisher}
  {Cambridge University Press},\ \bibinfo {year} {2014})\BibitemShut {NoStop}%
\bibitem [{\citenamefont {Ferreira}\ \emph {et~al.}(2017)\citenamefont
  {Ferreira}, \citenamefont {Macedo},\ and\ \citenamefont
  {Cardoso}}]{Ferreira:2017pth}%
  \BibitemOpen
  \bibfield  {author} {\bibinfo {author} {\bibfnamefont {M.~C.}\ \bibnamefont
  {Ferreira}}, \bibinfo {author} {\bibfnamefont {C.~F.~B.}\ \bibnamefont
  {Macedo}}, \ and\ \bibinfo {author} {\bibfnamefont {V.}~\bibnamefont
  {Cardoso}},\ }\href {\doibase 10.1103/PhysRevD.96.083017} {\bibfield
  {journal} {\bibinfo  {journal} {Phys. Rev. D}\ }\textbf {\bibinfo {volume}
  {96}},\ \bibinfo {pages} {083017} (\bibinfo {year} {2017})},\ \Eprint
  {http://arxiv.org/abs/1710.00830} {arXiv:1710.00830 [gr-qc]} \BibitemShut
  {NoStop}%
\bibitem [{\citenamefont {Detweiler}(1980)}]{Detweiler:1980uk}%
  \BibitemOpen
  \bibfield  {author} {\bibinfo {author} {\bibfnamefont {S.~L.}\ \bibnamefont
  {Detweiler}},\ }\href {\doibase 10.1103/PhysRevD.22.2323} {\bibfield
  {journal} {\bibinfo  {journal} {Phys. Rev. D}\ }\textbf {\bibinfo {volume}
  {22}},\ \bibinfo {pages} {2323} (\bibinfo {year} {1980})}\BibitemShut
  {NoStop}%
\bibitem [{\citenamefont {{Chandrasekhar}}(1943)}]{1943ApJ....97..255C}%
  \BibitemOpen
  \bibfield  {author} {\bibinfo {author} {\bibfnamefont {S.}~\bibnamefont
  {{Chandrasekhar}}},\ }\href {\doibase 10.1086/144517} {\bibfield  {journal}
  {\bibinfo  {journal} {\apj}\ }\textbf {\bibinfo {volume} {97}},\ \bibinfo
  {pages} {255} (\bibinfo {year} {1943})}\BibitemShut {NoStop}%
\bibitem [{\citenamefont {Liu}\ \emph {et~al.}(2014)\citenamefont {Liu},
  \citenamefont {Muñoz},\ and\ \citenamefont {Lai}}]{2014}%
  \BibitemOpen
  \bibfield  {author} {\bibinfo {author} {\bibfnamefont {B.}~\bibnamefont
  {Liu}}, \bibinfo {author} {\bibfnamefont {D.~J.}\ \bibnamefont {Muñoz}}, \
  and\ \bibinfo {author} {\bibfnamefont {D.}~\bibnamefont {Lai}},\ }\href
  {\doibase 10.1093/mnras/stu2396} {\bibfield  {journal} {\bibinfo  {journal}
  {Monthly Notices of the Royal Astronomical Society}\ }\textbf {\bibinfo
  {volume} {447}},\ \bibinfo {pages} {747–764} (\bibinfo {year}
  {2014})}\BibitemShut {NoStop}%
\bibitem [{\citenamefont {Su}\ \emph {et~al.}(2021)\citenamefont {Su},
  \citenamefont {Lai},\ and\ \citenamefont {Liu}}]{Su:2020vda}%
  \BibitemOpen
  \bibfield  {author} {\bibinfo {author} {\bibfnamefont {Y.}~\bibnamefont
  {Su}}, \bibinfo {author} {\bibfnamefont {D.}~\bibnamefont {Lai}}, \ and\
  \bibinfo {author} {\bibfnamefont {B.}~\bibnamefont {Liu}},\ }\href {\doibase
  10.1103/PhysRevD.103.063040} {\bibfield  {journal} {\bibinfo  {journal}
  {Phys. Rev. D}\ }\textbf {\bibinfo {volume} {103}},\ \bibinfo {pages}
  {063040} (\bibinfo {year} {2021})},\ \Eprint
  {http://arxiv.org/abs/2010.11951} {arXiv:2010.11951 [gr-qc]} \BibitemShut
  {NoStop}%
\bibitem [{\citenamefont {Crescini}\ \emph {et~al.}(2020)\citenamefont
  {Crescini} \emph {et~al.}}]{QUAX:2020adt}%
  \BibitemOpen
  \bibfield  {author} {\bibinfo {author} {\bibfnamefont {N.}~\bibnamefont
  {Crescini}} \emph {et~al.} (\bibinfo {collaboration} {QUAX}),\ }\href
  {\doibase 10.1103/PhysRevLett.124.171801} {\bibfield  {journal} {\bibinfo
  {journal} {Phys. Rev. Lett.}\ }\textbf {\bibinfo {volume} {124}},\ \bibinfo
  {pages} {171801} (\bibinfo {year} {2020})},\ \Eprint
  {http://arxiv.org/abs/2001.08940} {arXiv:2001.08940 [hep-ex]} \BibitemShut
  {NoStop}%
\bibitem [{\citenamefont {Gramolin}\ \emph {et~al.}(2021)\citenamefont
  {Gramolin}, \citenamefont {Aybas}, \citenamefont {Johnson}, \citenamefont
  {Adam},\ and\ \citenamefont {Sushkov}}]{Gramolin:2020ict}%
  \BibitemOpen
  \bibfield  {author} {\bibinfo {author} {\bibfnamefont {A.~V.}\ \bibnamefont
  {Gramolin}}, \bibinfo {author} {\bibfnamefont {D.}~\bibnamefont {Aybas}},
  \bibinfo {author} {\bibfnamefont {D.}~\bibnamefont {Johnson}}, \bibinfo
  {author} {\bibfnamefont {J.}~\bibnamefont {Adam}}, \ and\ \bibinfo {author}
  {\bibfnamefont {A.~O.}\ \bibnamefont {Sushkov}},\ }\href {\doibase
  10.1038/s41567-020-1006-6} {\bibfield  {journal} {\bibinfo  {journal} {Nature
  Phys.}\ }\textbf {\bibinfo {volume} {17}},\ \bibinfo {pages} {79} (\bibinfo
  {year} {2021})},\ \Eprint {http://arxiv.org/abs/2003.03348} {arXiv:2003.03348
  [hep-ex]} \BibitemShut {NoStop}%
\end{thebibliography}%
\end{document}